\documentclass[
twocolumn,
groupedaddress,
superscriptaddress,
amsmath,
amssymb,
amsfonts, 
amstext,
aps,
prl,
]{revtex4-2}

\usepackage{dcolumn}
\usepackage{bm}
\usepackage{soul}
\usepackage[flushleft]{threeparttable}
\usepackage{url}
\usepackage{mathrsfs}
\usepackage{braket}
\usepackage{multirow}
\usepackage{array}
\usepackage{booktabs}
\usepackage{float}
\usepackage{lineno}
\usepackage{hyperref}
\usepackage{mathtools}
\usepackage{xcolor}
\usepackage{enumitem}
\usepackage{tikz}

\newcommand*{\circled}[2][]{%
  \lower.7ex\hbox{\tikz\draw (0pt, 0pt) circle (.42em) node {\makebox[1em][c]{\footnotesize #2}};}%
  \ifx&#1&%
  \else
    \caption[#1]{#1}%
  \fi
}

\hypersetup{colorlinks=true, urlcolor=blue, citecolor=blue, linkcolor=blue, pdfborder={0 0 0}}

\newcommand{\eg}{\textit{e.g. }}
\begin{document}

\title{Hybrid-order topology in unconventional magnets of Eu-based Zintl compounds
with surface-dependent quantum geometry}

\author{Yufei Zhao}
\affiliation{Department of Condensed Matter Physics, Weizmann Institute of Science, Rehovot 7610001, Israel}
\author{Yiyang Jiang}
\affiliation{Department of Condensed Matter Physics, Weizmann Institute of Science, Rehovot 7610001, Israel}
\author{Hyeonhu Bae}
\affiliation{Department of Condensed Matter Physics, Weizmann Institute of Science, Rehovot 7610001, Israel}
\author{Kamal Das}
\affiliation{Department of Condensed Matter Physics, Weizmann Institute of Science, Rehovot 7610001, Israel}
\author{Yongkang Li}
\affiliation{Department of Condensed Matter Physics, Weizmann Institute of Science, Rehovot 7610001, Israel}
\author{Chao-Xing Liu}
\affiliation{Department of Physics, the Pennsylvania State University, University Park, PA 16802, USA}
\author{Binghai Yan}
\email{binghai.yan@weizmann.ac.il}
\affiliation{Department of Condensed Matter Physics, Weizmann Institute of Science, Rehovot 7610001, Israel}

\date{\today}

\begin{abstract}
The exploration of magnetic topological insulators is instrumental in exploring axion electrodynamics and intriguing transport phenomena, such as the quantum anomalous Hall effect. Here, we report that a family of magnetic compounds Eu$_{2n+1}$In$_{2}$(As,Sb)$_{2n+2}$ ($n=0,1,2$) exhibit both gapless Dirac surface states and chiral hinge modes. 
Such a hybrid-order topology hatches surface-dependent quantum geometry. 
By mapping the responses into real space, we demonstrate the existence of
chiral hinge modes along the $c$ direction, which originate from 
the half-quantized anomalous Hall effect on two gapped $ac$/$bc$ facets due to Berry curvature, while the unpinned Dirac surface states on the gapless $ab$ facet generate an intrinsic nonlinear anomalous Hall effect due to the quantum metric. 
When Eu$_{3}$In$_{2}$As$_{4}$ is polarized to the ferromagnetic phase by an external magnetic field, it becomes an ideal Weyl semimetal with a single pair of type-I Weyl points and no extra Fermi pocket. Our work predicts rich topological states sensitive to magnetic structures, quantum geometry-induced transport and topological superconductivity if proximitized with a superconductor.  

\end{abstract}

\maketitle

\section{Introduction}
\begin{figure*}[t]
    \centering
    \includegraphics[width=0.9\linewidth]{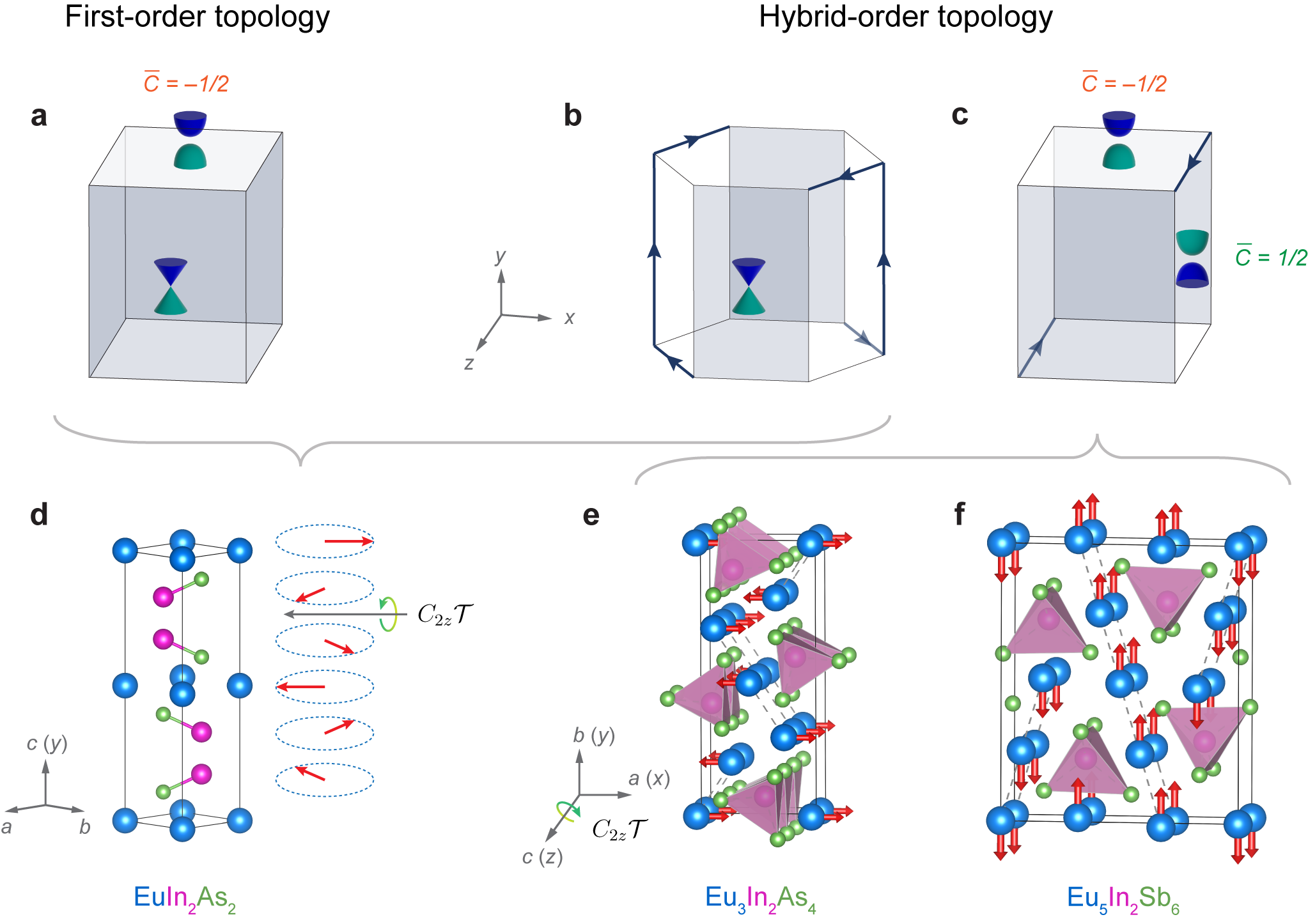}
    \caption{Symmetry protected boundary states, crystal and magnetic structures. \textbf{a} The first-order topology with gapless surfaces (dark) protected by $\tau_{y} \mathcal{T}$ symmetry, exemplified by MnBi$_{2}$Te$_{4}$. \textbf{b,c} The hybrid-order topology with unpinned Dirac surface (dark) and chiral hinge states (thick lines) under $C_{2z}$$\mathcal{T}$ symmetry, exemplified by Eu$_{2n+1}$In$_{2}$(As,Sb)$_{2n+2}$. \textbf{b} and \textbf{c} show two kinds of sample geometry. \textbf{d} EuIn$_{2}$As$_{2}$ (space group: $P6_{3}/mmc$): Broken helix spin order (in-$ab$-plane) below 16 K \cite{riberolles2021magnetic,soh2023understanding,donoway2023symmetrybreaking}. Six layers of Eu atoms (dashed ellipses) form a period. \textbf{e} Eu$_{3}$In$_{2}$As$_{4}$ ($Pnnm$): AFM order predicted by calculations. The thin dashed lines mark a set of sublattices. \textbf{f} Eu$_{5}$In$_{2}$Sb$_{6}$ ($Pbam$): Collinear magnetic order between 7 and 14 K \cite{rahn2023unusual,PhysRevB.109.014432}. Below 7 K, the spin order becomes coplanar but $C_{2z}\mathcal{T}$ symmetry still survives. For Eu$_{3}$In$_{2}$As$_{4}$ and Eu$_{5}$In$_{2}$Sb$_{6}$, the $C_{2z}\mathcal{T}$ axis passes through the cell center.}
    \label{fig1}
\end{figure*}

Topological quantum states of matter are usually characterized by unique boundary states following the bulk-boundary correspondence \cite{RevModPhys.82.3045, qi2011topological, tokura2019magnetic}. To be specific, a topological insulator (TI) exhibits two-dimensional (2D) Dirac-type surface states \cite{chen2009experimental, zhang2009topological} and a higher-order TI shows 1D hinge states \cite{hsieh2012topological, schindler2018higher, yue2019symmetry, doi:10.1126/sciadv.1602415}. Unlike nonmagnetic topological crystalline insulators (TCI) \cite{benalcazar2017quantized, hsieh2012topological, song2017d, schindler2018higher, PhysRevLett.123.186401}, surface/hinge states in a magnetic TCI are intimately related to magnetic structures \cite{yue2019symmetry}. Applying a magnetic field to flip or rotate spin moments is equivalent to reversing the mass signs of the gapped surface, thus switching the chirality of hinge states and the mapped 2D Chern number \cite{chang2013experimental, yue2019symmetry, zhang2020mobius, XuPhysRevLett.122.256402, doi:10.1126/science.aan5991}. Identifying the nature of topological invariants under a combined magnetic crystalline symmetry by the translation ($\tau$) or $n$-fold rotation ($C_{n}$) with time-reversal ($\mathcal{T}$), is evidently a central issue \cite{PhysRevB.85.165120, PhysRevB.91.161105, PhysRevLett.121.106403, PhysRevB.99.235125, PhysRevB.106.195144}. For instance, MnBi$_{2}$Te$_{4}$ is a $\tau \mathcal{T}$ protected antiferromagnetic (AFM) TI with gapless Dirac surface parallel to $\tau$ \cite{PhysRevB.81.245209, PhysRevLett.122.206401, li2019intrinsic}, namely a first-order topology depicted in Fig. \hyperref[fig1]{1a}. A hybrid-order topological phase ($e.g.$, Fig. \hyperref[fig1]{1b-c}), which exhibits both topological surface states and hinge states, is rare among intrinsic magnetic materials \cite{zhang2020mobius}.

Furthermore, a magnetic TCI with $\tau\mathcal{T}$ or $C_{n}\mathcal{T}$ symmetry can generate axion electrodynamics, described by the so-called quantized $\theta$ term \cite{PhysRevB.78.195424}, $S_{\theta} = \int dt d^{3}x \frac{\theta e^{2}}{4\pi^2 \hbar c} \bm{E} \cdot \bm{B}$, 
where $\bm{E}$ and $\bm{B}$ are external electric and magnetic fields, $\theta$ is the axion field defined in a 3D system \cite{PhysRevLett.102.146805, armitage2019matter, fang2012bulk, PhysRevB.81.245209, 10.1063/5.0038804, PhysRevLett.122.206401}. Axion insulators manifest a half-quantized anomalous Hall conductance (AHC) of $\pm \frac{e^{2}}{2h}$ on the gapped surface \cite{tokura2019magnetic,liu2020robust, PhysRevB.98.245117}, which comes from Berry curvature, $i.e.$ the imaginary part of the quantum geometry tensor \cite{provost1980riemannian, RevModPhys.82.1959}. When two adjacent gapped surfaces hold opposite masses, the surface quantum anomalous Hall effect (QAHE) is carried by the 1D chiral hinge mode \cite{gu2021spectral}. 
On the other side, the real part of the quantum geometry--quantum metric--was recently discovered to generate nonlinear anomalous Hall effect (NLAHE) in magnetic conductors \cite{PhysRevLett.127.277201, PhysRevLett.127.277202, wang2023quantum, gao2023quantum, PhysRevLett.132.026301}. We are inspired to ask whether metallic surface states can generate a 
surface NLAHE due to the quantum metric in a magnetic TCI. 

Recently, Eu-based Zintl compounds Eu$_{2n+1}$In$_{2}$(As,Sb)$_{2n+2}$ ($n=0,1,2$) emerged with unusual magnetic structures revealed by neutron diffraction, resonant elastic X-ray scattering, muon spin-rotation, or magnetotransport \cite{CHILDS2019120889, PhysRevB.101.205126,  PhysRevResearch.2.033342, riberolles2021magnetic, soh2023understanding, donoway2023symmetrybreaking, rosa2020colossal, crivillero2022surface, rahn2023unusual, PhysRevB.109.014432, haim324,jia2024discovery}. 
These materials can be magnetic TCI candidates due to the interplay of AFM order and strong spin-orbit coupling (SOC). 
In EuIn$_{2}$As$_{2}$, a pure helix order formed at 17.5 K evolves to a `broken helix order' below 16 K (Fig. \hyperref[fig1]{1d}) \cite{PhysRevB.101.205126,  PhysRevResearch.2.033342, riberolles2021magnetic, soh2023understanding, donoway2023symmetrybreaking}, although an earlier theory study assumed a collinear AFM order \cite{XuPhysRevLett.122.256402}.
In Eu$_{3}$In$_{2}$As$_{4}$, an AFM order was conformed below $5\sim6.5$ K in experiments \cite{haim324,jia2024discovery}. 
In Eu$_{5}$In$_{2}$As$_{6}$, a collinear AFM order formed at 14 K (Fig. \hyperref[fig1]{1f}) evolves to a complex coplanar order below 7 K accompanied with the anisotropic susceptibility \cite{rosa2020colossal,crivillero2022surface, rahn2023unusual, PhysRevB.109.014432}. 
Their magnetic structures of three compounds exhibit $C_2\mathcal{T}$ symmetry but break $\tau\mathcal{T}$ and $\mathcal{PT}$, different from the conventional AFM phase ($e.g.$, MnBi$_2$Te$_4$) but similar to `altermagnets', a collinear AFM order studied recently \cite{PhysRevX.12.021016, PhysRevX.12.031042, PhysRevX.12.040501, krempasky2024altermagnetic, zhu2024observation}.
Additionally, Eu$_{3}$In$_{2}$As$_{4}$ was recently grown from InAs nanowires in a mutual-exchange method \cite{haim324} and inherits the ideal proximity with a superconductor \cite{das2012zero}. Thus, this system will serve as a potential platform to study topological superconductivity due to the interplay between chiral hinge states/Dirac surface states and the $s$-wave superconductivity.

In this work, we predict that Eu$_{2n+1}$In$_{2}$(As,Sb)$_{2n+2}$ compounds possess hybrid-order topology characterized by the coexistence of
chiral hinge modes and unpinned Dirac surface states protected by a $C_{2}$$\mathcal{T}$-related topological invariant \cite{PhysRevB.91.161105, PhysRevLett.121.106403, PhysRevB.99.235125, PhysRevB.106.195144, PhysRevLett.121.106403, PhysRevB.99.235125}. Taking Eu$_{3}$In$_{2}$As$_{4}$ for an example, we demonstrate the axion insulator state with a surface QAHE. On the gapless surface, we find the NLAHE due to the quantum metric from Dirac states. In addition, the AFM bulk can exhibit the anomalous Hall effect despite vanishing net magnetization, because the magnetic lattice violates $\tau\mathcal{T}$ and $\mathcal{PT}$. 
If the magnetic order turns into ferromagnetic (FM) by applying a magnetic field, Eu$_{3}$In$_{2}$As$_{4}$ exhibits an ideal Weyl semimetal phase. We further derive a topological phase diagram based on a material-specific effective model to reveal possible topological states for these compounds in different magnetic orders.  
Finally, we discuss possible unique Majorana modes when the material is in proximity to an ordinary superconductor.

\section{Results}
\textbf{Magnetic order and bulk band structure. } 
We first focus on Eu$_{3}$In$_{2}$As$_{4}$ to demonstrate the hybrid-order topology and resultant properties. 
As illustrated in Fig. \hyperref[fig1]{1e}, the Zintl compound Eu$_{3}$In$_{2}$As$_{4}$ crystallizes in an orthorhombic lattice with the inversion symmetric space group $Pnnm$, including a two-fold rotation $C_{2z}$, and nonsymmorphic symmetries $\{ C_{2x,2y}\vert \tau \}$, where $\tau$ is the translation of one-half of a body diagonal $(\frac{a}{2}, \frac{b}{2}, \frac{c}{2})$. All atoms are located at $z=0$ and $z=0.5$ planes, and one In atom and four As atoms form a tetrahedron. The tetrahedrons ([In$_{2}$As$_{4}$]$^{6-}$) constitute a 1D chain by sharing their corners along the $c$ axis, between which the divalent Eu$^{2+}$ cations ($\mathbf{S} = \frac{7}{2}$) are dispersed. Three Eu atoms (linked by the dashed line) with the two nearest neighboring tetrahedrons form a sublattice. Within one unit cell, two sublattices are related by $\{ C_{2x,2y}\vert \tau \}$ operations.

The exact spin structure has thus far not been resolved in experiments, so we carry out the calculations to examine the magnetic ground state of Eu$_{3}$In$_{2}$As$_{4}$. Inspired by the collinear spin order of the sister compound Eu$_{5}$In$_{2}$As$_{6}$ \cite{rahn2023unusual, PhysRevB.109.014432}, we investigated several collinear magnetic structures (Supplementary Table III). We found an AFM phase with the easy axis in $ab$-plane (Fig. \hyperref[fig1]{1e}) is energetically favored, consistent with recent magnetometry measurements \cite{haim324, jia2024discovery}. The Eu$^{2+}$ cations exhibit FM coupling along the $c$-axis while two sublattices are AFM coupled. The in-$ab$-plane magnetic structure 
preserves $C_{2z}\mathcal{T}$ symmetry.

\begin{figure*}[t]
    \centering
    \includegraphics[width=0.9\textwidth]{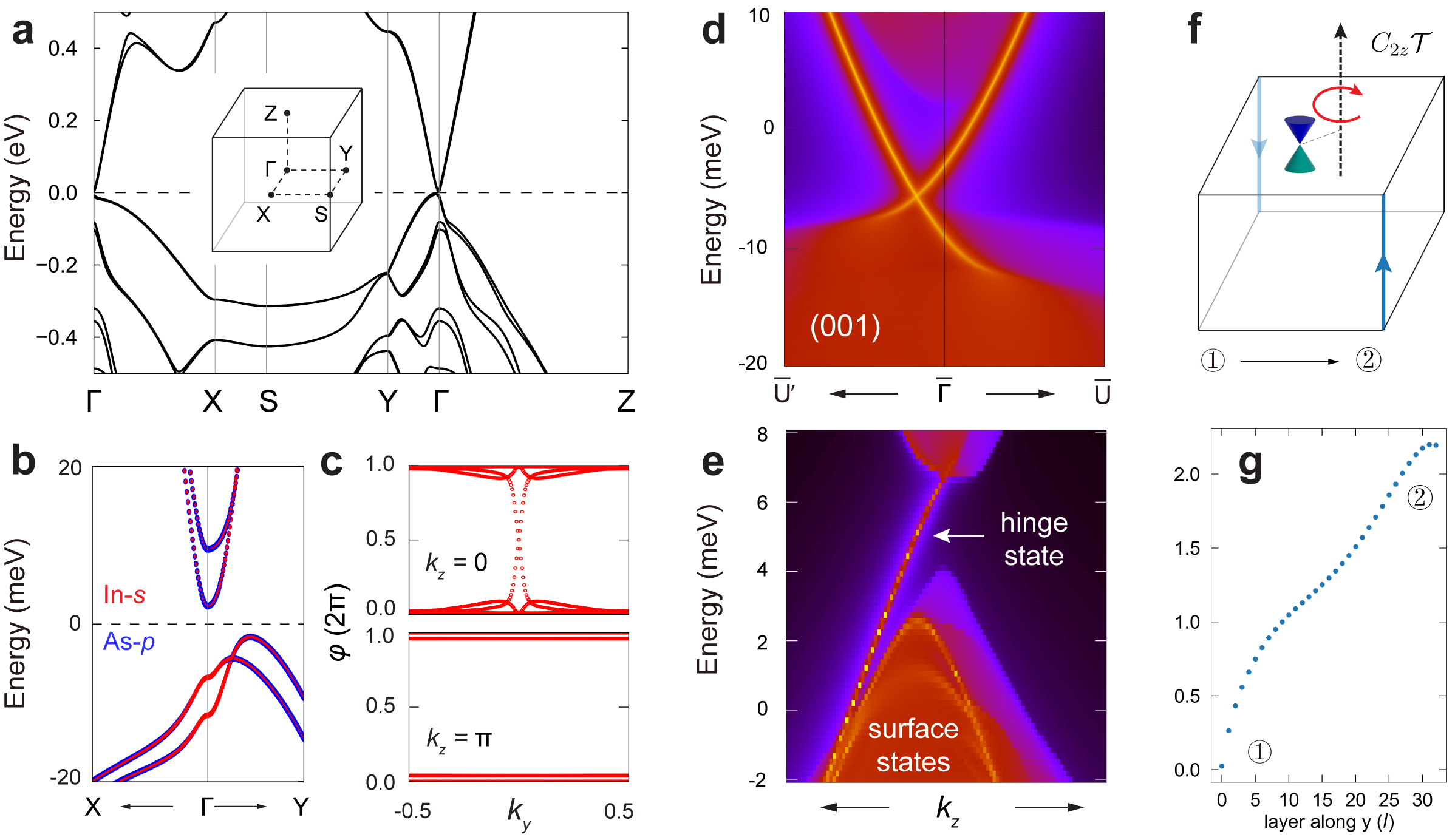}
    \caption{Bulk-surface-hinge correspondence in Eu$_{3}$In$_{2}$As$_{4}$. \textbf{a} Bulk band structure. \textbf{b} Inverted band gap with orbital weights.  \textbf{c} The 1D $k_{x}$-integrated Wilson bands on the $k_z=0/\pi$ planes. \textbf{d} (001) Surface states. $\overline{\mathrm{U}}$ is not a high-symmetry point. \textbf{e} The hinge state flowing along $c$-axis, calculated in a finite 34.5-unit-cell along $b$-axis ($\hat{y}$) and semi-infinite along $a$-axis ($\hat{x}$).
    \textbf{f} Schematic illustration of the unpinned Dirac point under $C_{2z}\mathcal{T}$.  Hinges \circled{1}/\circled{2} is adjacent by the $bc$ facet and $ac$ facet. \textbf{g} The intensity of the in-gap states (pointed in \textbf{e}) projected onto each layer (unit cell), showing a localization behavior at hinge \circled{2}. \#0 and \#35 represent two hinges.
    }
    \label{fig2}
\end{figure*}

The nonmagnetic phase gives a trivial semiconducting band structure while magnetic orders lead to nontrivial topology. The band structure of Eu$_{3}$In$_{2}$As$_{4}$ with an AFM$a$ order (easy axis along $a$) is shown in Fig. \hyperref[fig2]{2a}. Localized Eu-4$f$ orbitals appear below the Fermi energy between $-2.5$ eV and $-1.5$ eV. Without SOC, a trivial bandgap of 52 meV is found at $\Gamma$ point and spin splitting appears along $\Gamma-S$ (see Supplementary Fig. S9). The inclusion of SOC induces an inverted gap of nearly 10 meV at $\Gamma$ with band inversion and an indirect global gap of 5 meV (Fig. \hyperref[fig2]{2b}), which is larger than the energy scale of magnetic order temperature ($T_{N} \sim$ 5 K) or induced superconductivity gap. The spin double-degeneracy is lifted along $\Gamma-X$, $\Gamma-Y$, and $\Gamma-Z$, simialr to an altermanget, but remains along $X-S-Y$ because of antiunitary operators $\{ C_{2x}\vert \tau \}\mathcal{T}$ and $\{ M_{x}\vert \tau \}\mathcal{T}$. Details of symmetry analysis can be found in the Supplementary Sec. III.

\textbf{Surface and hinge states.} 
The antiunitary symmetry $C_{2z}\mathcal{T}$ helps understand the bulk-surface-hinge correspondence. Let us begin with a $\mathcal{T}$-preserved $\mathbb{Z}_2$ TI where all surfaces present Dirac surface states. For an orthorhombic crystal (Fig. \hyperref[fig1]{1c}), magnetism opens a surface gap on $bc$ and $ac$ facets, $i.e.$, introduces a mass term in the Dirac equation. 
Because they transform to each other by $C_{2z}\mathcal{T}$, two $bc$ facets (also two $ac$ facets) exhibit opposite masses.
For four hinges between $ac$ and $bc$ facets along the $c$ direction, we can always find two diagonal hinges at the domain boundary between surfaces with opposite signs of mass, leading to chiral hinge modes inside the surface gap. $C_{2z}\mathcal{T}$ symmetry is preserved on the $ab$ plane and leads to crystalline symmetry-protected unpinned Dirac surface states. The surface Dirac cone can only be pushed away from time-reversal-invariant momenta (\eg $\Bar{\Gamma}$) rather than gapped out. Two chiral hinge states terminate at gapless (001) surfaces while the equilibrium current persists, different from an ordinary TI or TCI that shows only surface or hinge states. Consequently, three surface Chern numbers ($\overline{\mathbb{C}} = \pm \frac{1}{2}, 0$) can be realized on different surfaces, offering a road for designing the direction-selective QAHE and quantized circular dichroism effect \cite{PhysRevLett.123.247401}. 

Such a magnetic TCI is equivalent to a 3D strong Stiefel-Whitney insulator with vanishing Berry curvature 
due to $C_{2z}\mathcal{T}$ symmetry\cite{PhysRevLett.121.106403, PhysRevB.99.235125, pan2022two, guo2022quadrupole}. We can define a $\mathbb{Z}_{2}$-type topological invariant $\Delta_{{\mathrm{3D}}} = w_{2}(\pi) - w_{2}(0)$, where $w_{2}(k_{z})$ is the second Stiefel-Whitney number defined on a 2D plane. The first Stiefel-Whitney number $w_{1}$ is equivalent to the quantized Berry phase and thus $w_{1} = 0$ in our case. A nontrivial $w_{2}$ characterizes a double band inversion, which is well-defined only when $w_{1} = 0$. We can calculate $w_{2}$ by tracing the Wilson loop spectra $\varphi$, where $w_{2}$ is the number of crossings at $\varphi = \pi$ mod 2. Fig. \hyperref[fig2]{2c} indicates AFM$a$ phase as a 3D strong Stiefel-Whitney insulator ($\Delta_{{\mathrm{3D}}} = 1$) with $w_{2}(0) = 1$ and $w_{2}(\pi) = 0$.
The Stiefel-Whitney insulator can exhibit an odd number of surface Dirac cones \cite{PhysRevB.91.161105, PhysRevB.99.235125}. 

We perform the Green's function calculations using Wannier functions extracted from DFT calculations to verify the surface and hinge states. A single gapless Dirac cone exists on the (001) surface, shown in Fig. \hyperref[fig2]{2d}. Because there is no additional symmetry to constrain the Dirac point besides $C_{2z}\mathcal{T}$, 
the Dirac point is unpinned from $\Bar{\Gamma}$ or any high-symmetry line. The Dirac point positions can be continuously shifted by synchronously rotating or slightly titling the magnetic moments on the $ab$ plane (Supplementary Fig. S5), for which $C_{2z}\mathcal{T}$ is maintained. In this case, the in-plane magnetic moments act as a 
vector potential perturbation
that keeps the gapless nature of the surface Dirac cone.

The hinge modes of two adjacent side facets are analyzed by building a half-infinite slab model, containing 34 unit cells plus half a unit cell (34.5) along the $b$ axis, which respects $C_{2z}\mathcal{T}$ and is thick enough to suppress inter-facet and inter-hinge interactions. The obtained spectrum includes the states from bulk, one (100) surface, two (010) surfaces, and two hinges. In Fig. \hyperref[fig2]{2e}, an up-moving in-gap state clearly emerges as we expect. By projecting it into each layer, we find that the spectral weight is mainly distributed at hinge $\circled{2}$ (Fig. \hyperref[fig2]{2g}), confirming it as a hinge mode in the real space. These exotic boundary states attract us to next consider what kind of transport signals or physical quantity they carry on the surface.


\textbf{Surface quantum geometry.} To resolve a local response in real space, we can introduce the non-Abelian quantum geometric tensor in terms of two band ($n,m$), and depose it as \cite{PhysRevB.98.115108},
\begin{equation}
    \mathcal{Q}^{ij}_{nm} \equiv \mathcal{A}_{nm}^{i} \mathcal{A}_{mn}^{j} = g^{ij}_{nm} - \frac{i}{2}\Omega^{ij}_{nm},
    \label{eq1}
\end{equation}
where $\mathcal{A}_{nm}^{i} = i\braket{n|\partial_{k_{i}}|m}$ is the Berry connection, $g_{nm}^{ij}$ and $\Omega_{vv^{\prime}}^{ij}$ are the quantum metric and Berry curvature.
Then the local AHC in a slab is identified by the layer-resolved Chern number $\sigma_{\mathrm{slab}}^{ij}(l) = \frac{e^2}{h}\mathbb{C}(l) $ \cite{PhysRevB.98.245117}
\begin{equation}
    \sigma_{\mathrm{slab}}^{ij}(l) = \frac{e^{2}}{\pi h} \mathrm{Im} \sum_{n=1}^{N_c} \sum_{m=N_{c}+1}^{\infty} \int_{\bm{k}} \frac{\braket{n|\hat{P_{l}} \partial_{k_{i}} \mathcal{H}|m} \braket{m|\partial_{k_{j}} \mathcal{H}|n}}{(\varepsilon_{n} - \varepsilon_{m})^{2}},
    \label{eq2}
\end{equation}
where $N_{c}$ is the number of occupied bands, $\hat{P_{l}} = \ket{l} \bra{l}$ is the projector onto $l$-th layer. 

\begin{figure*}
    \centering
    \includegraphics[width=0.75\linewidth]{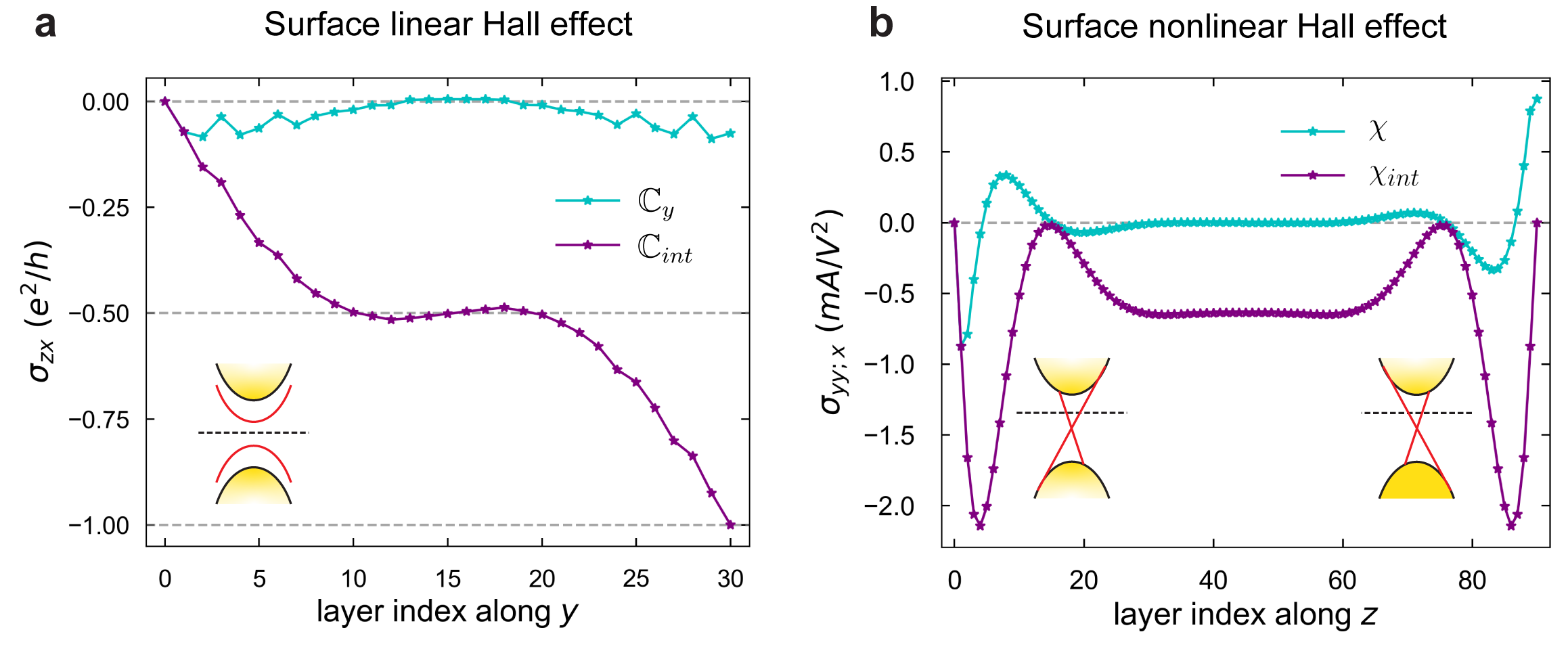}
    \caption{Surface anomalous Hall effect. \textbf{a} Local Chern number and integrated Hall conductance as a function of layers in a 29.5-unit-cell slab. \textbf{b} Intrinsic second-order Hall conductivity from the quantum metric dipole in a 90-unit-cell slab. The Fermi level lies in the surface gap in \textbf{a} and cuts through the surface Dirac cone in \textbf{b}. }
    \label{fig3}
\end{figure*}

By constructing a slab along the $y$ direction, $\mathbb{C}_{y}(l)$ and $\mathbb{C}_{int} = \sum _{l} \mathbb{C}_{y}(l)$ in Fig. \hyperref[fig3]{3a} show a saturation behavior of the half-quantized surface AHC. Such a local signal is known to be the direct evidence of the axion insulator \cite{PhysRevB.98.245117}. The first ten layers are enough to stabilize $\mathbb{C}_{int}$ at 0.4976 (corresponds to the penetration length of the hinge state), while the internal layers from \#10 to \#20 (bulk) tame the oscillation and do not contribute to $\sigma^{zx}$ of the entire slab. Since the spin moments of the top and bottom layers hold the same direction, the half-quantized surface AHC from the first and last ten layers together gives rise to a $\mathbb{C}=-1$ Chern insulator.

Besides the surface AHC, it is worth noting that $C_{2z}\mathcal{T}$ does not exclude all the bulk Hall tensors. When the system is doped to be a conductor, a finite AHC $\sigma^{zx}_{\mathrm{bulk}}$ will also emerge despite the compensated magnetization (Supplementary Fig. S10), offering a signature of unusual antiferromagnet or altermagnet.

On the $C_{2z}\mathcal{T}$-preserved $ab$ facet, the linear AHC $\sigma^{xy}$ vanishes but the second-order Hall conductivity is permitted ($\sigma^{ij;k}=\mathcal{J}^{k}/E^{i}E^{j}$) \cite{PhysRevLett.132.026301}. In a similar vein, the layer-resolved second-order Hall conductivity can be given by
\begin{widetext}
\begin{equation}
    \sigma_{\mathrm{slab}}^{ii;j}(l) = \frac{e^{3}}{\hbar} \mathrm{Re} \sum_{n, m \neq n}^{\infty} \int_{\bm{k}} \frac{ 2 \bra{n} \hat{P_{l}} \partial_{k_{j}} \mathcal{H} \ket{n} \bra{n}\partial_{k_{i}} \mathcal{H} \ket{m} \bra{m} \partial_{k_{i}} \mathcal{H} \ket{n} - \bra{n} \hat{P_{l}} \partial_{k_{i}} \mathcal{H} \ket{n} \bra{n} \partial_{k_{i}} \mathcal{H} \ket{m} \bra{m} \partial_{k_{j}} \mathcal{H} \ket{n} }{(\varepsilon_{n} - \varepsilon_{m})^{3}} \frac{\partial f_{n} }{\partial \varepsilon_{n}},
    \label{eq3}
\end{equation}
\end{widetext}
The Eq. \ref{eq3} includes the Fermi surface contribution from the scattering time-independent quantum metric dipole ($\chi^{\mathrm{QMD}}$). In Fig. \hyperref[fig3]{3b}, a localized intrinsic NLAHE is found in a slab stacked along the $c$ direction. Due to the inversion symmetry, the total second-order conductivity ($\chi_{int}$) is zero and the Dirac cone from the top and bottom surfaces are oppositely titled. Since $C_{2z}\mathcal{T}$ does not restrict the shape of the Dirac cone, an in-plane nonlinear response $\sigma^{yy;x}$ will appear on the surface. Breaking inversion with different chemical potentials for the top and bottom surface states is expected to induce a net NLAHE in the thin film, which can be probed in transport experiments. In addition, both the surface AHE and surface NLAHE discussed above can also be measured by the nonlocal transport, for example, by using a device configuration proposed in Ref.~\cite{PhysRevB.103.L241409}.

\begin{figure*}[t]
    \centering
    \includegraphics[width=0.85\linewidth]{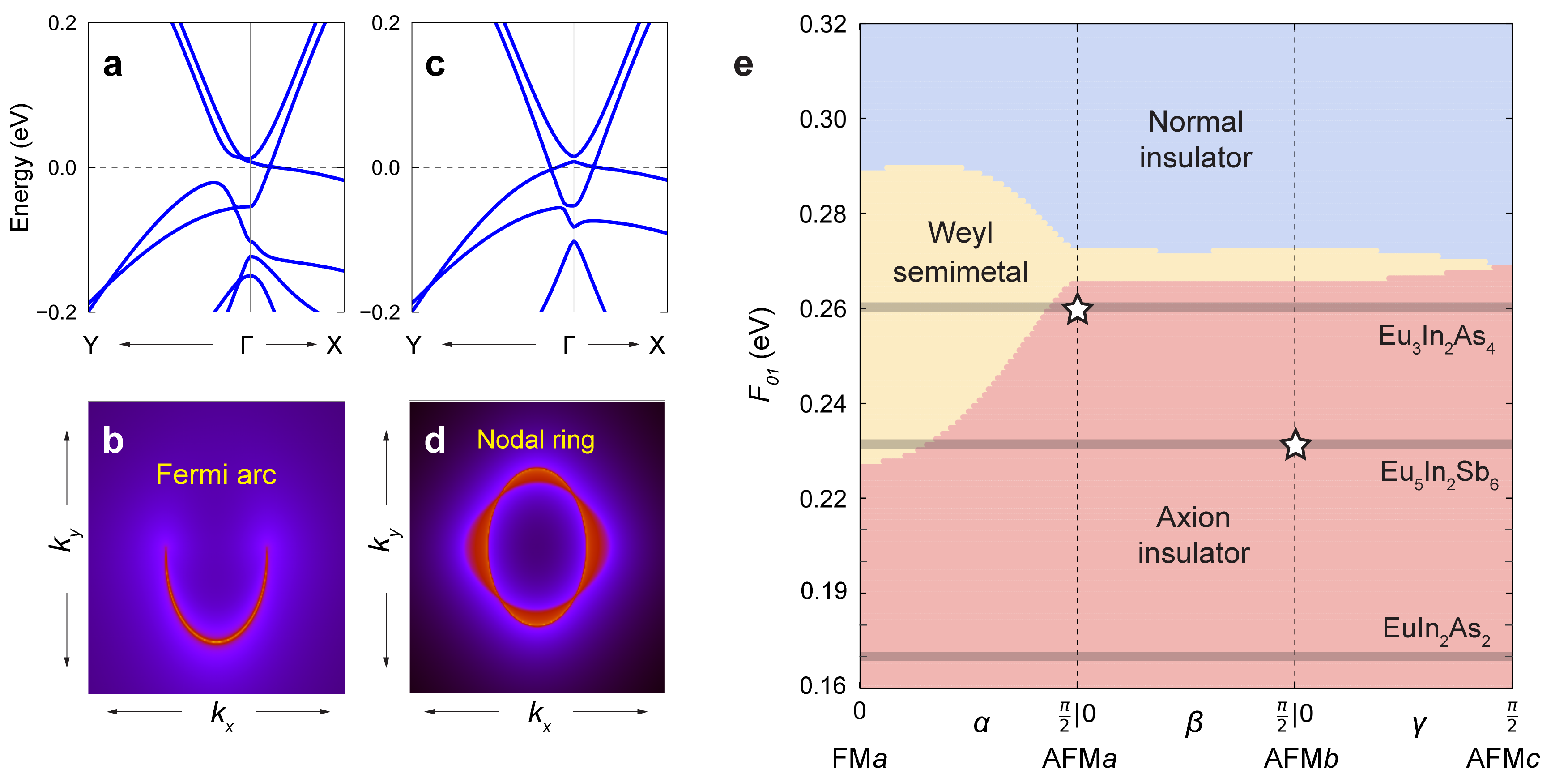}
    \caption{Multiple magnetic topological phases in Zintl compounds. \textbf{a,c} Band structure near $\Gamma$ point of the FM$a$ and FM$c$ phases. \textbf{b,d} Projected (001) Fermi surface. The Fermi arc connects two Weyl points in \textbf{b}. Symmetry $M_{z}$ protects a Weyl nodal ring in \textbf{c}. \textbf{e} Topological phase diagram in terms of SOC strength $F_{01}$ and magnetic orders. Pentagrams along the gray line represent the ground states in Fig. 2\textbf{a} and Supplementary Fig. S11. }
    \label{fig4}
\end{figure*}

\textbf{Weyl semimetal phase.} When all the spin moments align along the $a$ axis under an external field (FM$a$, $\textbf{m} \parallel a$), we find two type-I Weyl points with opposite chiralities along $\Gamma-X$ connected by inversion symmetry in Fig. \hyperref[fig4]{4a,b}. In the projected (001) Fermi surface, two Weyl nodes separated 0.074 \AA$^{-1}$ along $k_{x}$ are connected by a surface Fermi arc. In the FM$b$ phase (Supplementary Fig. S7), the Weyl nodes are shifted to locate along the $\Gamma-Y$ direction. Thus, the FM$a/b$ phase is an ideal Weyl semimetal, which exhibits a single pair of type-I Weyl points without extra trivial Fermi pockets at the Fermi energy. In the FM$c$ phase, the mirror $M_{z}$ will protect a nodal ring on the $k_{z} = 0$ plane and drumhead-like surface states, as shown in Fig. \hyperref[fig4]{4c,d}.

\textbf{Effective model and topological phase diagram.} 
We build a general topological phase diagram by constructing a $k \cdot p$ model using the invariant theory \cite{PhysRevB.82.045122}, to more intuitively understand the role of magnetic order in the topological band structure. Since $\{ C_{2y} \vert \tau \}$ is a key symmetry to relate two sublattices and the AFM order, three sets of Pauli matrices should be introduced to describe this system:
\begin{itemize}
    \item spin $\sigma$: $\ket{\frac{1}{2}}$ and $\ket{-\frac{1}{2}}$
    \item orbital $\tau$: $\ket{\mathrm{In} \ s}$ and $\ket{\mathrm{As} \ p_{z}}$
    \item sublattice $\zeta$: $\ket{A}$ and $\ket{B}$
\end{itemize}
and the basis are labeled by $\ket{\Lambda_{\alpha}, \uparrow(\downarrow)}$, where $\Lambda$ = In, As; $\alpha$ = $A,B$; $\uparrow(\downarrow)$ = $\frac{1}{2}(-\frac{1}{2})$. According to the band representation (parity $\pm$) and wavefunction at $\Gamma$, we can have the orbital basis $\ket{\mathrm{In}^{+}_{A} \ s},\ket{\mathrm{In}^{+}_{B} \ s},\ket{\mathrm{As}^{-}_{A} \ p_{z}},\ket{\mathrm{As}^{-}_{B} \ p_{z}}$. The symmetry operators are given by $\hat{\mathcal{T}} = i\sigma_{2}\tau_{0}\zeta_{0}\mathcal{K}$, inversion $\hat{\mathcal{I}} = \sigma_{0}\tau_{3}\zeta_{0}$, $\hat{C}_{2z} = -i\sigma_{3}\tau_{0}\zeta_{0}$ and $\{ \hat{C}_{2y}\vert \tau \}  = -i\sigma_{2}\tau_{3}\zeta_{1}$, where $\mathcal{K}$ is the complex conjugate operator. Then we can list the irreducible representation (IR) of each matrix in Supplementary Sec. I. Based on that, the effective Hamiltonian is described by

\begin{widetext}
    \begin{equation}
    \mathcal{H} = \mathcal{H}_{\mathrm{NM}} + \mathcal{H}_{ex} = 
    \begin{pmatrix}
        \mathcal{H}_{A} & \mathcal{H}_{AB} \\
        \mathcal{H}_{AB}^{\dagger} & \mathcal{H}_{B} \\
    \end{pmatrix}
    +
    \begin{pmatrix}
        \mathcal{H}_{\mathrm{M},A} & \mathcal{H}_{\mathrm{M},AB} \\
        \mathcal{H}_{\mathrm{M},AB}^{\dagger} & \mathcal{H}_{\mathrm{M},B} \\
    \end{pmatrix}
    \end{equation}
    \begin{equation}
    \begin{aligned}
    \mathcal{H}_{A/B} = &(\pm A_{1}k_{x} + A_{2}k_{y}) \sigma_{1}\tau_{1} + (B_{1}k_{x} \pm B_{2} k_{y})\sigma_{2}\tau_{1} \pm C_{1}k_{z} \sigma_{3}\tau_{1}\\
    &+ D_{1}k_{z} \sigma_{0}\tau_{2} + E_{01} \sigma_{0}\tau_{0} + (F_{01} + F_{1}k_{x}^{2} + F_{2}k_{y}^{2} + F_{3}k_{z}^2 \pm F_{4}k_{x}k_{y}) \sigma_{0}\tau_{3},
    \end{aligned}
    \end{equation}
    \begin{equation}
    \begin{aligned}
    \mathcal{H}_{AB} = & A_{3}k_{y}\sigma_{1}\tau_{1} + B_{3}k_{x}\sigma_{2}\tau_{1} + D_{2}k_{z}\sigma_{0}\tau_{2} + E_{02} \sigma_{0}\tau_{0} + F_{02} \sigma_{0}\tau_{3}\\
    & - i G_{0} \sigma_{3}\tau_{0} - i L_{0} \sigma_{3}\tau_{3} - iM_{1}k_{x}\sigma_{1}\tau_{2} - iN_{1}k_{y}\sigma_{2}\tau_{2} - iO_{1}k_{z}\sigma_{3}\tau_{2},
    \end{aligned}
    \end{equation}
\end{widetext}
where $F_{01}$ controls the bandgap so it can be approximately treated as the inverse of the effective SOC strength. Experimentally, $F_{01}$ is tuned by pnictide doping or pressure \cite{varnava2022engineering}. The other material-dependent parameters are chosen to make dispersion close to $\Gamma$ qualitatively or semi-quantitatively consistent with Fig. \hyperref[fig2]{2b}.

We next consider the exchange coupling terms $\mathcal{H}_{ex}$, namely $\mathcal{T}$-odd terms when introducing spin orders. Note that the AFM$a$ phase shares the same IR with FM$b$ (also for AFM$b$ and FM$a$), so both the on-site terms $\sigma_{1}\tau_{0}\zeta_{3}$ and $\sigma_{2}\tau_{0}\zeta_{0}$ will couple with IR $\Gamma_{4}^{+}$. Generally, $\bm{\mathrm{m}}\cdot\bm{\sigma}$ could be taken to characterize the magnetization of one sublattice and thus to distinguish them.
Take the AFM$a$ phase as an example, we have
\begin{equation}
    \begin{aligned}
    \mathcal{H}_{\mathrm{AFM}a,A/B} = &\pm \Delta_{x1}\sigma_{1}\tau_{0} \pm \Delta_{x2}\sigma_{1}\tau_{3},
    \end{aligned}
\end{equation}
\begin{equation}
    \begin{aligned}
    \mathcal{H}_{\mathrm{AFM}a,AB} = \Delta_{3}\sigma_{2}\tau_{0} + \Delta_{4}\sigma_{2}\tau_{3}.
    \end{aligned}
\end{equation}
Similarly, the exchange terms for the FM$a$, AFM$b$, and AFM$c$ phases can be obtained, respectively. Additionally, the intermediate phases between them can be expressed as a linear combination.

By numerically solving the model, we construct a $\mathcal{T}$-broken phase diagram (Fig. \hyperref[fig5]{5}) as functions of $F_{01}$ and $\alpha, \beta, \gamma$. We find Eu$_{3}$In$_{2}$As$_{4}$ can be very well mapped onto the grey horizontal line with $F_{01}=0.26$ eV. As an intermediate region between two insulating phases, the chance of being a Weyl semimetal is intimately related to the magnetic configuration. Within the FM order, the Zeeman effect results in a large semimetallic region. Then it shrinks to a fairly narrow range in the AFM$a/b$ or tilted phase, and eventually vanishes in the AFM$c$ phase. The axion insulator phase is quite robust against the perturbation on orientations of the AFM order in Eu$_{3}$In$_{2}$As$_{4}$. The model parameters and verification are provided in the Supplementary Sec. III.

\section{Discussion} 
\textbf{Materials.}
We emphasize the results above also apply to EuIn$_{2}$As$_{2}$ and Eu$_{5}$In$_{2}$Sb$_{6}$, which are fitted in the phase diagram by only considering the anticrossing energy. To observe hinge states, EuIn$_{2}$As$_{2}$ is a bit special for the presence of both $C_{2z}\mathcal{T}$ and $C_{2x}\mathcal{T}$ symmetries by the `broken helix order' \cite{riberolles2021magnetic}. When the sample is in the orthorhombic geometry (Fig. \hyperref[fig1]{1a}), the system presents a first-order topology with the only gapped (001) surface. In the hexagonal geometry (Fig. \hyperref[fig1]{1b}), $C_{2x}\mathcal{T}$ axis passing through the center of two hinges reverts the gap masses of neighboring facets so that the hinge current can appear along the $c$ direction ($\hat{y}$). Eu$_{5}$In$_{2}$Sb$_{6}$ exhibits the same boundary states as Eu$_{3}$In$_{2}$As$_{4}$ because spin moments are aligned in a collinear or coplanar way to present only $C_{2z}\mathcal{T}$ \cite{PhysRevB.109.014432}.

The hybrid-order topology could also emerge in systems summarized in Table \ref{table1} and hatches different surface-dependent quantum geometry. 
Besides $\chi^{\mathrm{QMD}}$, the second-order Hall conductivity also includes contributions from the Drude weight $\chi^{\mathrm{D}}$ (nongeometric) and Berry curvature dipole $\chi^{\mathrm{BCD}}$ (imaginary part of the quantum geometry) \cite{PhysRevLett.132.026301}. On the $C_{2}\mathcal{T}$-preserved surfaces, the surface quantum geometry only contains $\chi^{\mathrm{QMD}}$. In other systems, the surface NLAHE tensors allowed are elaborated below and in Supplementary Sec. IV.
First, in $\mathcal{T}$-preserved TCIs, $C_{2,4,6}$ symmetry eliminates all the surface $\chi^{ii;j}$ tensors while mirror/glide planes only permits $\chi^{\mathrm{BCD}}$.
Second, in $\mathcal{T}$-broken TCIs, the mirror/glide symmetry accommodates both $\chi^{\mathrm{BCD}}$ and $\chi^{\mathrm{QMD}}$. However, it is fragile against magnetic perturbation and thus hard to protect surface/hinge states. Overall, the surface NLAHE is driven by surface states rather than hinge states.

\begin{table}[t]
\caption{\label{table2} Emergence of a hybrid-order topology classified from symmetries. The details are explained in Supplementary Sec. IV. Here, the mirror plane ($M$) must go through both the crystal surface and hinges ($e.g.$ SnTe). The glide symmetry ($G$) is restricted to exist only on one set of surfaces [$e.g.$ (100) surface of the canted AFM MnBi$_{2}$Te$_{4}$ \cite{zhang2020mobius}]. When more than one following symmetries exist, the hybrid-order topology may be quenched to the first-order. 
}
\renewcommand\arraystretch{1.5}
\begin{ruledtabular}
\begin{tabular}{p{1cm}cp{2cm}p{3cm}}
Hinge states  & \multirow{2}{*}{$\mathcal{T}$} &  Symmetries on the surface & Surface QAHE and Surface NLAHE\\ \hline
\multirow{2}{*}{Helical}  & \multirow{2}{*}{$\checkmark$} &  $C_{2}$, $C_{4}$, $C_{6}$ &  $-$ \\
& & $M$ \cite{hsieh2012topological}, $G$ \cite{doi:10.1126/sciadv.1602415} & $\chi^{\mathrm{BCD}}$ \\ \hline
\multirow{3}{*}{Chiral}   & \multirow{3}{*}{$\times$} &  $C_{4}\mathcal{T}$ & $\overline{\mathbb{C}}$ \\
 & & $C_{2}\mathcal{T}$, $C_{6}\mathcal{T}$ & $\overline{\mathbb{C}}$, $\chi^{\mathrm{D}}$, $\chi^{\mathrm{QMD}}$ \\
& & $M$, $G$ \cite{zhang2020mobius} & $\overline{\mathbb{C}}$, $\chi^{\mathrm{D}}$, $\chi^{\mathrm{BCD}}$, $\chi^{\mathrm{QMD}}$
\end{tabular}
\end{ruledtabular}
\label{table1}
\end{table}

\begin{figure}[t]
    \centering
    \includegraphics[width=\linewidth]{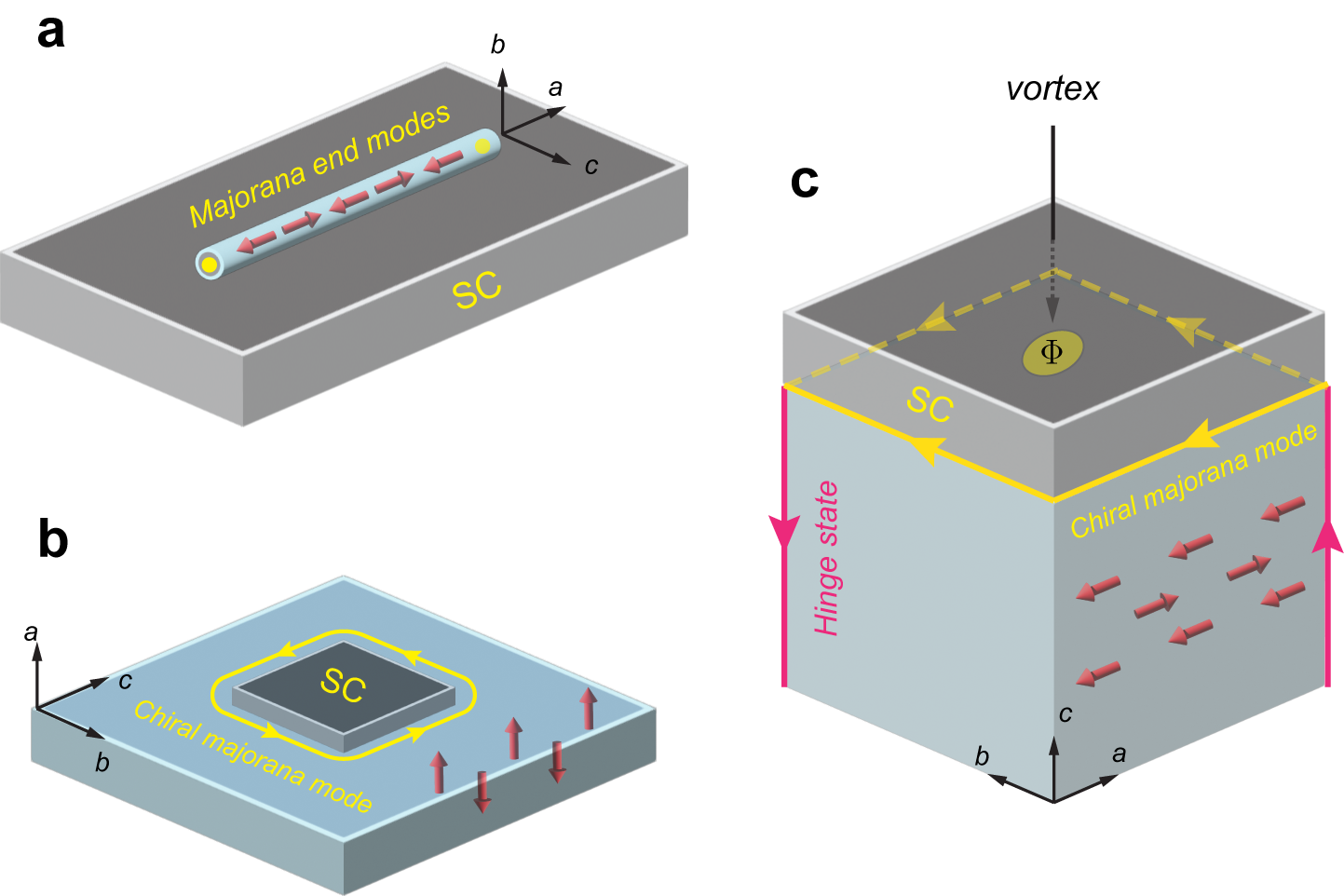}
    \caption{Schmatics of the axion insulator Eu$_{3}$In$_{2}$As$_{4}$ (cyan-blue) in proximity to an $s$-wave superconductor (SC) (grey). (a) The nanowire configuration. (b) The SC proximity to the gapped surface, $e.g.$, the $ac/bc$ surfaces in AFM$a$ phase. (c) The SC proximity to the gapless surface, $e.g.$, the $ab$ surface in AFM$a$ phase. One chiral hinge state (pink lines) splits into two chiral Majorana edge modes, providing an interferometer configuration. Yellow lines (dots) represent the chiral Majorana modes (Majorana end modes). The red arrows represent spin moments in the axion insulator.
    }
    \label{fig5}
\end{figure}

\textbf{Topological superconductivity.} Eu$_{3}$In$_{2}$As$_{4}$ nanowires were recently grown from InAs nanowires by a topotaxial mutual-exchange method~\cite{haim324} and can inherit good interface quality to the ordinary superconductor ($e.g.$, Al) \cite{das2012zero}.
Because of the intrinsic magnetism, strong SOC effect, and topological surface states, Eu$_{3}$In$_{2}$As$_{4}$ paves an appealing avenue to explore Majorana physics, as illustrated in Fig. \hyperref[fig5]{5}.
The intrinsic magnetism and SOC together lift the spin degeneracy and an odd number of Fermi surfaces can be realized at an accessible chemical potential, which is the precondition to design structure with Majorana \cite{PhysRevLett.100.096407, ghorashi2023altermagnetic}. 
($i$) In a nanowire configuration \cite{Lutchyn2010,Oreg2010}, the finite size effect gaps the hinge and surface states, leading to nondegenerate topology-driven nanowire bands. In proximity to an $s$-wave superconductor, localized zero-energy Majorana modes will arise at two ends. Compared to earlier proposals \cite{ PhysRevB.84.201105, alicea2012new}, we do not require an external magnetic field to break time-reversal symmetry here. 
($ii$) On the gapped surface ($e.g.$, $ac/bc$ facet), one can deposit a superconductor island and expect chiral Majorana edge states around the island, if the superconductor's gap $\Delta$ overcomes the magnetization-induced surface gap $h$. Using Fu and Kane's proposal \cite{PhysRevLett.100.096407, tokura2019magnetic}, we expect a stable nontrivial pairing by covering a superconductor on $ac$ or $bc$ surfaces (criterion $\vert h \vert<\Delta$ when the Fermi energy lies inside the gap). In this case, 1D chiral Majorana modes appear at the interface between the axion insulator's surface and a superconductor region \cite{PhysRevLett.102.216404}.
If the Fermi energy crosses an odd number of Dirac bands (metallic regime), similar chiral Majorana modes also appear in this case. 
($iii$) On the gapless surface ($e.g.$, $ab$ facet in AFM$a$), covering a superconductor on the whole surface may lead to co-propagating 1D chiral Majorana edge modes. Two Majorana modes merge into one chiral electronic state at the hinge \cite{PhysRevB.92.064520}.  As illustrated by Fig. \hyperref[fig5]{5c}, two chiral Majorana modes and two chiral hinge states naturally form a Mach–Zehnder interferometer, in which the conductance between two chiral hinge modes depends on the parity of the number of enclosed vortices in the superconductor region ~\cite{PhysRevLett.102.216404, Fu2009}. 

In summary, through first-principle calculations and $k \cdot p $ model, we reveal the coexistence of the unpinned Dirac surface and chiral hinge states in Eu$_{2n+1}$In$_{2}$(As,Sb)$_{2n+2}$.  Established by $C_{2}\mathcal{T}$ symmetry, such a hybrid-order topology also leads to a surface-dependent quantum geometry. Our findings on the nonlinear transport suggest new fingerprint evidence of magnetic axion insulators and other $\mathcal{T}$-preserved TCIs. Since the single crystals and nanowires of Eu$_{3}$In$_{2}$As$_{4}$ are readily synthesized, angle-resolved photoemission spectroscopy (ARPES), scanning tunneling spectroscopy (STS) and other techniques are called to detect these topological surface states, hinge states, and Weyl points/Fermi arcs and explore their coupling to magnetism and superconductivity.

~

\textbf{Note added.} After we posted the manuscript, an experiment work was posted to report a pair of Weyl nodes in the ferromagnetic phase under an external magnetic field \cite{jia2024discovery}, consistent with our theory.

\section{Methods}
\textbf{Density-functional theory.}
Density-functional theory (DFT) calculations are performed with the Vienna Ab initio Simulation Package (\textsc{vasp}) using the projector-augmented wave method \cite{PhysRevB.54.11169, PhysRevB.59.1758, PhysRevB.50.17953}. The Exchange-correlation functional based on generalized gradient approximation (GGA) parameterized by Perdew-Burke-Ernzerhof (PBE) is adopted \cite{PhysRevLett.77.3865}. The kinetic energy cutoff of the plane-wave basis is set to 400 eV. The Brillouin zone integration is performed by using 8×5×10 $\Gamma$-centered $\bm{k}$ point mesh. To treat the correlation effect of localized 4$f$ electrons of Eu, the DFT+$U$ method by Dudarev et al. is employed \cite{PhysRevB.57.1505}. The details about the DFT functional choice, Hubbard $U$, and the Green's function can be found in Supplementary Sec. II.

\textbf{Calculations of the layer-resolved Hall conductivity.}
We construct the maximally localized Wannier functions of In-5$s$ orbital and As-4$p$ orbitals by using \textsc{wannier90} package \cite{PhysRevB.65.035109}. The local Chern number (geometric part of the local AHC) is calculated by using a 121×121 $k$-mesh \cite{PhysRevB.98.245117}. The nonlinear surface Hall conductivity is performed on the $k \cdot p$ Hamiltonian using an 800×800 $k$-mesh. The complete formula of the second-order nonlinear conductivity reads as
\begin{equation}
\begin{aligned}
    \sigma_{ij;k} &= \chi_{ij;k}^{\mathrm{D}} + \chi_{ij;k}^{\mathrm{BCD}} + \chi_{ij;k}^{\mathrm{QMD}} \\
    &= -\frac{e^3 \tau^{2}}{\hbar^{3}} \sum_{n} \int_{\bm{k}} f_{n} \partial_{k_{i}}\partial_{k_{j}}\partial_{k_{k}} \varepsilon_{n} \\
    & ~~~ - \frac{e^3 \tau}{\hbar^{2}} \sum_{n} \int_{\bm{k}} f_{n} (\partial_{k_{i}} \Omega_{n}^{jk} + \partial_{k_{j}} \Omega_{n}^{ik}) \\
    & ~~~ - \frac{e^3}{\hbar} \sum_{n} \int_{\bm{k}} f_{n} \left( 2\partial_{k_{k}} \mathcal{G}_{\bm{k}}^{ij} - \frac{1}{2}(\partial_{k_{i}} \mathcal{G}_{n}^{jk} + \partial_{k_{j}} \mathcal{G}_{n}^{ik}) \right),
\end{aligned}
\end{equation}
where $\Omega_{n} = \sum_{m \neq n} \Omega_{nm}$ is the Berry curvature, and $ \mathcal{G}_{n} = \sum_{m \neq n} 2g_{nm} / (\varepsilon_{n} - \varepsilon_{m}) $ is the band-normalized quantum metric.

\section{Data availability}
All the data that support the findings of this study are available within the paper and Supplementary Information. Additional data are available from the corresponding authors upon reasonable request.

\section{Acknowledgments}
We are inspired by a recent experiment work by Haim Beidenkopf and Hadas Shtrikman. We thank helpful discussions with Xi Dai, Mingqiang Gu, Hengxin Tan and Zhiqiang Mao. C.-X. L. acknowledges the support of NSF Grant No. (DMR-2241327). B.Y. acknowledges the financial support by the European Research Council (ERC Consolidator Grant ``NonlinearTopo'', No. 815869) and the ISF - Personal Research Grant	(No. 2932/21). 

\section{Author contributions}
B.Y. conceived and supervised the project. Y.Z., and H.B. performed DFT calculations. Y.Z., Y.J., K.D carried out the surface quantum geometry anlysis. Y.Z. and C.-X. L carried out the model Hamiltonian. The paper was written by Y.Z. and B.Y. All authors contributed to
the discussion.

\section{Competing interests}
The authors declare no competing interests.


\begin{thebibliography}{79}%
\makeatletter
\providecommand \@ifxundefined [1]{%
 \@ifx{#1\undefined}
}%
\providecommand \@ifnum [1]{%
 \ifnum #1\expandafter \@firstoftwo
 \else \expandafter \@secondoftwo
 \fi
}%
\providecommand \@ifx [1]{%
 \ifx #1\expandafter \@firstoftwo
 \else \expandafter \@secondoftwo
 \fi
}%
\providecommand \natexlab [1]{#1}%
\providecommand \enquote  [1]{``#1''}%
\providecommand \bibnamefont  [1]{#1}%
\providecommand \bibfnamefont [1]{#1}%
\providecommand \citenamefont [1]{#1}%
\providecommand \href@noop [0]{\@secondoftwo}%
\providecommand \href [0]{\begingroup \@sanitize@url \@href}%
\providecommand \@href[1]{\@@startlink{#1}\@@href}%
\providecommand \@@href[1]{\endgroup#1\@@endlink}%
\providecommand \@sanitize@url [0]{\catcode `\\12\catcode `\$12\catcode
  `\&12\catcode `\#12\catcode `\^12\catcode `\_12\catcode `\%12\relax}%
\providecommand \@@startlink[1]{}%
\providecommand \@@endlink[0]{}%
\providecommand \url  [0]{\begingroup\@sanitize@url \@url }%
\providecommand \@url [1]{\endgroup\@href {#1}{\urlprefix }}%
\providecommand \urlprefix  [0]{URL }%
\providecommand \Eprint [0]{\href }%
\providecommand \doibase [0]{https://doi.org/}%
\providecommand \selectlanguage [0]{\@gobble}%
\providecommand \bibinfo  [0]{\@secondoftwo}%
\providecommand \bibfield  [0]{\@secondoftwo}%
\providecommand \translation [1]{[#1]}%
\providecommand \BibitemOpen [0]{}%
\providecommand \bibitemStop [0]{}%
\providecommand \bibitemNoStop [0]{.\EOS\space}%
\providecommand \EOS [0]{\spacefactor3000\relax}%
\providecommand \BibitemShut  [1]{\csname bibitem#1\endcsname}%
\let\auto@bib@innerbib\@empty
\bibitem [{\citenamefont {Hasan}\ and\ \citenamefont
  {Kane}(2010)}]{RevModPhys.82.3045}%
  \BibitemOpen
  \bibfield  {author} {\bibinfo {author} {\bibfnamefont {M.~Z.}\ \bibnamefont
  {Hasan}}\ and\ \bibinfo {author} {\bibfnamefont {C.~L.}\ \bibnamefont
  {Kane}},\ }\bibfield  {title} {\bibinfo {title} {Colloquium: Topological
  insulators},\ }\href {https://doi.org/10.1103/RevModPhys.82.3045} {\bibfield
  {journal} {\bibinfo  {journal} {Rev. Mod. Phys.}\ }\textbf {\bibinfo {volume}
  {82}},\ \bibinfo {pages} {3045} (\bibinfo {year} {2010})}\BibitemShut
  {NoStop}%
\bibitem [{\citenamefont {Qi}\ and\ \citenamefont
  {Zhang}(2011)}]{qi2011topological}%
  \BibitemOpen
  \bibfield  {author} {\bibinfo {author} {\bibfnamefont {X.-L.}\ \bibnamefont
  {Qi}}\ and\ \bibinfo {author} {\bibfnamefont {S.-C.}\ \bibnamefont {Zhang}},\
  }\bibfield  {title} {\bibinfo {title} {Topological insulators and
  superconductors},\ }\href@noop {} {\bibfield  {journal} {\bibinfo  {journal}
  {Reviews of Modern Physics}\ }\textbf {\bibinfo {volume} {83}},\ \bibinfo
  {pages} {1057} (\bibinfo {year} {2011})}\BibitemShut {NoStop}%
\bibitem [{\citenamefont {Tokura}\ \emph {et~al.}(2019)\citenamefont {Tokura},
  \citenamefont {Yasuda},\ and\ \citenamefont
  {Tsukazaki}}]{tokura2019magnetic}%
  \BibitemOpen
  \bibfield  {author} {\bibinfo {author} {\bibfnamefont {Y.}~\bibnamefont
  {Tokura}}, \bibinfo {author} {\bibfnamefont {K.}~\bibnamefont {Yasuda}},\
  and\ \bibinfo {author} {\bibfnamefont {A.}~\bibnamefont {Tsukazaki}},\
  }\bibfield  {title} {\bibinfo {title} {Magnetic topological insulators},\
  }\href@noop {} {\bibfield  {journal} {\bibinfo  {journal} {Nature Reviews
  Physics}\ }\textbf {\bibinfo {volume} {1}},\ \bibinfo {pages} {126} (\bibinfo
  {year} {2019})}\BibitemShut {NoStop}%
\bibitem [{\citenamefont {Chen}\ \emph {et~al.}(2009)\citenamefont {Chen},
  \citenamefont {Analytis}, \citenamefont {Chu}, \citenamefont {Liu},
  \citenamefont {Mo}, \citenamefont {Qi}, \citenamefont {Zhang}, \citenamefont
  {Lu}, \citenamefont {Dai}, \citenamefont {Fang} \emph
  {et~al.}}]{chen2009experimental}%
  \BibitemOpen
  \bibfield  {author} {\bibinfo {author} {\bibfnamefont {Y.}~\bibnamefont
  {Chen}}, \bibinfo {author} {\bibfnamefont {J.~G.}\ \bibnamefont {Analytis}},
  \bibinfo {author} {\bibfnamefont {J.-H.}\ \bibnamefont {Chu}}, \bibinfo
  {author} {\bibfnamefont {Z.}~\bibnamefont {Liu}}, \bibinfo {author}
  {\bibfnamefont {S.-K.}\ \bibnamefont {Mo}}, \bibinfo {author} {\bibfnamefont
  {X.-L.}\ \bibnamefont {Qi}}, \bibinfo {author} {\bibfnamefont
  {H.}~\bibnamefont {Zhang}}, \bibinfo {author} {\bibfnamefont
  {D.}~\bibnamefont {Lu}}, \bibinfo {author} {\bibfnamefont {X.}~\bibnamefont
  {Dai}}, \bibinfo {author} {\bibfnamefont {Z.}~\bibnamefont {Fang}}, \emph
  {et~al.},\ }\bibfield  {title} {\bibinfo {title} {Experimental realization of
  a three-dimensional topological insulator, bi2te3},\ }\href@noop {}
  {\bibfield  {journal} {\bibinfo  {journal} {science}\ }\textbf {\bibinfo
  {volume} {325}},\ \bibinfo {pages} {178} (\bibinfo {year}
  {2009})}\BibitemShut {NoStop}%
\bibitem [{\citenamefont {Zhang}\ \emph {et~al.}(2009)\citenamefont {Zhang},
  \citenamefont {Liu}, \citenamefont {Qi}, \citenamefont {Dai}, \citenamefont
  {Fang},\ and\ \citenamefont {Zhang}}]{zhang2009topological}%
  \BibitemOpen
  \bibfield  {author} {\bibinfo {author} {\bibfnamefont {H.}~\bibnamefont
  {Zhang}}, \bibinfo {author} {\bibfnamefont {C.-X.}\ \bibnamefont {Liu}},
  \bibinfo {author} {\bibfnamefont {X.-L.}\ \bibnamefont {Qi}}, \bibinfo
  {author} {\bibfnamefont {X.}~\bibnamefont {Dai}}, \bibinfo {author}
  {\bibfnamefont {Z.}~\bibnamefont {Fang}},\ and\ \bibinfo {author}
  {\bibfnamefont {S.-C.}\ \bibnamefont {Zhang}},\ }\bibfield  {title} {\bibinfo
  {title} {Topological insulators in bi2se3, bi2te3 and sb2te3 with a single
  dirac cone on the surface},\ }\href@noop {} {\bibfield  {journal} {\bibinfo
  {journal} {Nature physics}\ }\textbf {\bibinfo {volume} {5}},\ \bibinfo
  {pages} {438} (\bibinfo {year} {2009})}\BibitemShut {NoStop}%
\bibitem [{\citenamefont {Hsieh}\ \emph {et~al.}(2012)\citenamefont {Hsieh},
  \citenamefont {Lin}, \citenamefont {Liu}, \citenamefont {Duan}, \citenamefont
  {Bansil},\ and\ \citenamefont {Fu}}]{hsieh2012topological}%
  \BibitemOpen
  \bibfield  {author} {\bibinfo {author} {\bibfnamefont {T.~H.}\ \bibnamefont
  {Hsieh}}, \bibinfo {author} {\bibfnamefont {H.}~\bibnamefont {Lin}}, \bibinfo
  {author} {\bibfnamefont {J.}~\bibnamefont {Liu}}, \bibinfo {author}
  {\bibfnamefont {W.}~\bibnamefont {Duan}}, \bibinfo {author} {\bibfnamefont
  {A.}~\bibnamefont {Bansil}},\ and\ \bibinfo {author} {\bibfnamefont
  {L.}~\bibnamefont {Fu}},\ }\bibfield  {title} {\bibinfo {title} {Topological
  crystalline insulators in the snte material class},\ }\href@noop {}
  {\bibfield  {journal} {\bibinfo  {journal} {Nature communications}\ }\textbf
  {\bibinfo {volume} {3}},\ \bibinfo {pages} {982} (\bibinfo {year}
  {2012})}\BibitemShut {NoStop}%
\bibitem [{\citenamefont {Schindler}\ \emph {et~al.}(2018)\citenamefont
  {Schindler}, \citenamefont {Wang}, \citenamefont {Vergniory}, \citenamefont
  {Cook}, \citenamefont {Murani}, \citenamefont {Sengupta}, \citenamefont
  {Kasumov}, \citenamefont {Deblock}, \citenamefont {Jeon}, \citenamefont
  {Drozdov} \emph {et~al.}}]{schindler2018higher}%
  \BibitemOpen
  \bibfield  {author} {\bibinfo {author} {\bibfnamefont {F.}~\bibnamefont
  {Schindler}}, \bibinfo {author} {\bibfnamefont {Z.}~\bibnamefont {Wang}},
  \bibinfo {author} {\bibfnamefont {M.~G.}\ \bibnamefont {Vergniory}}, \bibinfo
  {author} {\bibfnamefont {A.~M.}\ \bibnamefont {Cook}}, \bibinfo {author}
  {\bibfnamefont {A.}~\bibnamefont {Murani}}, \bibinfo {author} {\bibfnamefont
  {S.}~\bibnamefont {Sengupta}}, \bibinfo {author} {\bibfnamefont {A.~Y.}\
  \bibnamefont {Kasumov}}, \bibinfo {author} {\bibfnamefont {R.}~\bibnamefont
  {Deblock}}, \bibinfo {author} {\bibfnamefont {S.}~\bibnamefont {Jeon}},
  \bibinfo {author} {\bibfnamefont {I.}~\bibnamefont {Drozdov}}, \emph
  {et~al.},\ }\bibfield  {title} {\bibinfo {title} {Higher-order topology in
  bismuth},\ }\href@noop {} {\bibfield  {journal} {\bibinfo  {journal} {Nature
  physics}\ }\textbf {\bibinfo {volume} {14}},\ \bibinfo {pages} {918}
  (\bibinfo {year} {2018})}\BibitemShut {NoStop}%
\bibitem [{\citenamefont {Yue}\ \emph {et~al.}(2019)\citenamefont {Yue},
  \citenamefont {Xu}, \citenamefont {Song}, \citenamefont {Weng}, \citenamefont
  {Lu}, \citenamefont {Fang},\ and\ \citenamefont {Dai}}]{yue2019symmetry}%
  \BibitemOpen
  \bibfield  {author} {\bibinfo {author} {\bibfnamefont {C.}~\bibnamefont
  {Yue}}, \bibinfo {author} {\bibfnamefont {Y.}~\bibnamefont {Xu}}, \bibinfo
  {author} {\bibfnamefont {Z.}~\bibnamefont {Song}}, \bibinfo {author}
  {\bibfnamefont {H.}~\bibnamefont {Weng}}, \bibinfo {author} {\bibfnamefont
  {Y.-M.}\ \bibnamefont {Lu}}, \bibinfo {author} {\bibfnamefont
  {C.}~\bibnamefont {Fang}},\ and\ \bibinfo {author} {\bibfnamefont
  {X.}~\bibnamefont {Dai}},\ }\bibfield  {title} {\bibinfo {title}
  {Symmetry-enforced chiral hinge states and surface quantum anomalous hall
  effect in the magnetic axion insulator bi$_{2-x}$sm$_{x}$se3},\ }\href@noop
  {} {\bibfield  {journal} {\bibinfo  {journal} {Nature Physics}\ }\textbf
  {\bibinfo {volume} {15}},\ \bibinfo {pages} {577} (\bibinfo {year}
  {2019})}\BibitemShut {NoStop}%
\bibitem [{\citenamefont {Ma}\ \emph {et~al.}(2017)\citenamefont {Ma},
  \citenamefont {Yi}, \citenamefont {Lv}, \citenamefont {Wang}, \citenamefont
  {Nie}, \citenamefont {Wang}, \citenamefont {Kong}, \citenamefont {Huang},
  \citenamefont {Richard}, \citenamefont {Zhang}, \citenamefont {Yaji},
  \citenamefont {Kuroda}, \citenamefont {Shin}, \citenamefont {Weng},
  \citenamefont {Bernevig}, \citenamefont {Shi}, \citenamefont {Qian},\ and\
  \citenamefont {Ding}}]{doi:10.1126/sciadv.1602415}%
  \BibitemOpen
  \bibfield  {author} {\bibinfo {author} {\bibfnamefont {J.}~\bibnamefont
  {Ma}}, \bibinfo {author} {\bibfnamefont {C.}~\bibnamefont {Yi}}, \bibinfo
  {author} {\bibfnamefont {B.}~\bibnamefont {Lv}}, \bibinfo {author}
  {\bibfnamefont {Z.}~\bibnamefont {Wang}}, \bibinfo {author} {\bibfnamefont
  {S.}~\bibnamefont {Nie}}, \bibinfo {author} {\bibfnamefont {L.}~\bibnamefont
  {Wang}}, \bibinfo {author} {\bibfnamefont {L.}~\bibnamefont {Kong}}, \bibinfo
  {author} {\bibfnamefont {Y.}~\bibnamefont {Huang}}, \bibinfo {author}
  {\bibfnamefont {P.}~\bibnamefont {Richard}}, \bibinfo {author} {\bibfnamefont
  {P.}~\bibnamefont {Zhang}}, \bibinfo {author} {\bibfnamefont
  {K.}~\bibnamefont {Yaji}}, \bibinfo {author} {\bibfnamefont {K.}~\bibnamefont
  {Kuroda}}, \bibinfo {author} {\bibfnamefont {S.}~\bibnamefont {Shin}},
  \bibinfo {author} {\bibfnamefont {H.}~\bibnamefont {Weng}}, \bibinfo {author}
  {\bibfnamefont {B.~A.}\ \bibnamefont {Bernevig}}, \bibinfo {author}
  {\bibfnamefont {Y.}~\bibnamefont {Shi}}, \bibinfo {author} {\bibfnamefont
  {T.}~\bibnamefont {Qian}},\ and\ \bibinfo {author} {\bibfnamefont
  {H.}~\bibnamefont {Ding}},\ }\bibfield  {title} {\bibinfo {title}
  {Experimental evidence of hourglass fermion in the candidate nonsymmorphic
  topological insulator khgsb},\ }\href
  {https://doi.org/10.1126/sciadv.1602415} {\bibfield  {journal} {\bibinfo
  {journal} {Science Advances}\ }\textbf {\bibinfo {volume} {3}},\ \bibinfo
  {pages} {e1602415} (\bibinfo {year} {2017})},\ \Eprint
  {https://arxiv.org/abs/https://www.science.org/doi/pdf/10.1126/sciadv.1602415}
  {https://www.science.org/doi/pdf/10.1126/sciadv.1602415} \BibitemShut
  {NoStop}%
\bibitem [{\citenamefont {Benalcazar}\ \emph {et~al.}(2017)\citenamefont
  {Benalcazar}, \citenamefont {Bernevig},\ and\ \citenamefont
  {Hughes}}]{benalcazar2017quantized}%
  \BibitemOpen
  \bibfield  {author} {\bibinfo {author} {\bibfnamefont {W.~A.}\ \bibnamefont
  {Benalcazar}}, \bibinfo {author} {\bibfnamefont {B.~A.}\ \bibnamefont
  {Bernevig}},\ and\ \bibinfo {author} {\bibfnamefont {T.~L.}\ \bibnamefont
  {Hughes}},\ }\bibfield  {title} {\bibinfo {title} {Quantized electric
  multipole insulators},\ }\href@noop {} {\bibfield  {journal} {\bibinfo
  {journal} {Science}\ }\textbf {\bibinfo {volume} {357}},\ \bibinfo {pages}
  {61} (\bibinfo {year} {2017})}\BibitemShut {NoStop}%
\bibitem [{\citenamefont {Song}\ \emph {et~al.}(2017)\citenamefont {Song},
  \citenamefont {Fang},\ and\ \citenamefont {Fang}}]{song2017d}%
  \BibitemOpen
  \bibfield  {author} {\bibinfo {author} {\bibfnamefont {Z.}~\bibnamefont
  {Song}}, \bibinfo {author} {\bibfnamefont {Z.}~\bibnamefont {Fang}},\ and\
  \bibinfo {author} {\bibfnamefont {C.}~\bibnamefont {Fang}},\ }\bibfield
  {title} {\bibinfo {title} {(d- 2)-dimensional edge states of rotation
  symmetry protected topological states},\ }\href@noop {} {\bibfield  {journal}
  {\bibinfo  {journal} {Physical review letters}\ }\textbf {\bibinfo {volume}
  {119}},\ \bibinfo {pages} {246402} (\bibinfo {year} {2017})}\BibitemShut
  {NoStop}%
\bibitem [{\citenamefont {Wang}\ \emph {et~al.}(2019)\citenamefont {Wang},
  \citenamefont {Wieder}, \citenamefont {Li}, \citenamefont {Yan},\ and\
  \citenamefont {Bernevig}}]{PhysRevLett.123.186401}%
  \BibitemOpen
  \bibfield  {author} {\bibinfo {author} {\bibfnamefont {Z.}~\bibnamefont
  {Wang}}, \bibinfo {author} {\bibfnamefont {B.~J.}\ \bibnamefont {Wieder}},
  \bibinfo {author} {\bibfnamefont {J.}~\bibnamefont {Li}}, \bibinfo {author}
  {\bibfnamefont {B.}~\bibnamefont {Yan}},\ and\ \bibinfo {author}
  {\bibfnamefont {B.~A.}\ \bibnamefont {Bernevig}},\ }\bibfield  {title}
  {\bibinfo {title} {Higher-order topology, monopole nodal lines, and the
  origin of large fermi arcs in transition metal dichalcogenides
  $x{\mathrm{te}}_{2}$ ($x=\mathrm{Mo},\mathrm{W}$)},\ }\href
  {https://doi.org/10.1103/PhysRevLett.123.186401} {\bibfield  {journal}
  {\bibinfo  {journal} {Phys. Rev. Lett.}\ }\textbf {\bibinfo {volume} {123}},\
  \bibinfo {pages} {186401} (\bibinfo {year} {2019})}\BibitemShut {NoStop}%
\bibitem [{\citenamefont {Chang}\ \emph {et~al.}(2013)\citenamefont {Chang},
  \citenamefont {Zhang}, \citenamefont {Feng}, \citenamefont {Shen},
  \citenamefont {Zhang}, \citenamefont {Guo}, \citenamefont {Li}, \citenamefont
  {Ou}, \citenamefont {Wei}, \citenamefont {Wang} \emph
  {et~al.}}]{chang2013experimental}%
  \BibitemOpen
  \bibfield  {author} {\bibinfo {author} {\bibfnamefont {C.-Z.}\ \bibnamefont
  {Chang}}, \bibinfo {author} {\bibfnamefont {J.}~\bibnamefont {Zhang}},
  \bibinfo {author} {\bibfnamefont {X.}~\bibnamefont {Feng}}, \bibinfo {author}
  {\bibfnamefont {J.}~\bibnamefont {Shen}}, \bibinfo {author} {\bibfnamefont
  {Z.}~\bibnamefont {Zhang}}, \bibinfo {author} {\bibfnamefont
  {M.}~\bibnamefont {Guo}}, \bibinfo {author} {\bibfnamefont {K.}~\bibnamefont
  {Li}}, \bibinfo {author} {\bibfnamefont {Y.}~\bibnamefont {Ou}}, \bibinfo
  {author} {\bibfnamefont {P.}~\bibnamefont {Wei}}, \bibinfo {author}
  {\bibfnamefont {L.-L.}\ \bibnamefont {Wang}}, \emph {et~al.},\ }\bibfield
  {title} {\bibinfo {title} {Experimental observation of the quantum anomalous
  hall effect in a magnetic topological insulator},\ }\href@noop {} {\bibfield
  {journal} {\bibinfo  {journal} {Science}\ }\textbf {\bibinfo {volume}
  {340}},\ \bibinfo {pages} {167} (\bibinfo {year} {2013})}\BibitemShut
  {NoStop}%
\bibitem [{\citenamefont {Zhang}\ \emph
  {et~al.}(2020{\natexlab{a}})\citenamefont {Zhang}, \citenamefont {Wu},\ and\
  \citenamefont {Sarma}}]{zhang2020mobius}%
  \BibitemOpen
  \bibfield  {author} {\bibinfo {author} {\bibfnamefont {R.-X.}\ \bibnamefont
  {Zhang}}, \bibinfo {author} {\bibfnamefont {F.}~\bibnamefont {Wu}},\ and\
  \bibinfo {author} {\bibfnamefont {S.~D.}\ \bibnamefont {Sarma}},\ }\bibfield
  {title} {\bibinfo {title} {M{\"o}bius insulator and higher-order topology in
  mnbi$_{2n}$te$_{3n+1}$},\ }\href@noop {} {\bibfield  {journal} {\bibinfo
  {journal} {Physical review letters}\ }\textbf {\bibinfo {volume} {124}},\
  \bibinfo {pages} {136407} (\bibinfo {year} {2020}{\natexlab{a}})}\BibitemShut
  {NoStop}%
\bibitem [{\citenamefont {Xu}\ \emph {et~al.}(2019)\citenamefont {Xu},
  \citenamefont {Song}, \citenamefont {Wang}, \citenamefont {Weng},\ and\
  \citenamefont {Dai}}]{XuPhysRevLett.122.256402}%
  \BibitemOpen
  \bibfield  {author} {\bibinfo {author} {\bibfnamefont {Y.}~\bibnamefont
  {Xu}}, \bibinfo {author} {\bibfnamefont {Z.}~\bibnamefont {Song}}, \bibinfo
  {author} {\bibfnamefont {Z.}~\bibnamefont {Wang}}, \bibinfo {author}
  {\bibfnamefont {H.}~\bibnamefont {Weng}},\ and\ \bibinfo {author}
  {\bibfnamefont {X.}~\bibnamefont {Dai}},\ }\bibfield  {title} {\bibinfo
  {title} {Higher-order topology of the axion insulator
  ${\mathrm{euin}}_{2}{\mathrm{as}}_{2}$},\ }\href
  {https://doi.org/10.1103/PhysRevLett.122.256402} {\bibfield  {journal}
  {\bibinfo  {journal} {Phys. Rev. Lett.}\ }\textbf {\bibinfo {volume} {122}},\
  \bibinfo {pages} {256402} (\bibinfo {year} {2019})}\BibitemShut {NoStop}%
\bibitem [{\citenamefont {Yasuda}\ \emph {et~al.}(2017)\citenamefont {Yasuda},
  \citenamefont {Mogi}, \citenamefont {Yoshimi}, \citenamefont {Tsukazaki},
  \citenamefont {Takahashi}, \citenamefont {Kawasaki}, \citenamefont {Kagawa},\
  and\ \citenamefont {Tokura}}]{doi:10.1126/science.aan5991}%
  \BibitemOpen
  \bibfield  {author} {\bibinfo {author} {\bibfnamefont {K.}~\bibnamefont
  {Yasuda}}, \bibinfo {author} {\bibfnamefont {M.}~\bibnamefont {Mogi}},
  \bibinfo {author} {\bibfnamefont {R.}~\bibnamefont {Yoshimi}}, \bibinfo
  {author} {\bibfnamefont {A.}~\bibnamefont {Tsukazaki}}, \bibinfo {author}
  {\bibfnamefont {K.~S.}\ \bibnamefont {Takahashi}}, \bibinfo {author}
  {\bibfnamefont {M.}~\bibnamefont {Kawasaki}}, \bibinfo {author}
  {\bibfnamefont {F.}~\bibnamefont {Kagawa}},\ and\ \bibinfo {author}
  {\bibfnamefont {Y.}~\bibnamefont {Tokura}},\ }\bibfield  {title} {\bibinfo
  {title} {Quantized chiral edge conduction on domain walls of a magnetic
  topological insulator},\ }\href {https://doi.org/10.1126/science.aan5991}
  {\bibfield  {journal} {\bibinfo  {journal} {Science}\ }\textbf {\bibinfo
  {volume} {358}},\ \bibinfo {pages} {1311} (\bibinfo {year}
  {2017})}\BibitemShut {NoStop}%
\bibitem [{\citenamefont {Turner}\ \emph {et~al.}(2012)\citenamefont {Turner},
  \citenamefont {Zhang}, \citenamefont {Mong},\ and\ \citenamefont
  {Vishwanath}}]{PhysRevB.85.165120}%
  \BibitemOpen
  \bibfield  {author} {\bibinfo {author} {\bibfnamefont {A.~M.}\ \bibnamefont
  {Turner}}, \bibinfo {author} {\bibfnamefont {Y.}~\bibnamefont {Zhang}},
  \bibinfo {author} {\bibfnamefont {R.~S.~K.}\ \bibnamefont {Mong}},\ and\
  \bibinfo {author} {\bibfnamefont {A.}~\bibnamefont {Vishwanath}},\ }\bibfield
   {title} {\bibinfo {title} {Quantized response and topology of magnetic
  insulators with inversion symmetry},\ }\href
  {https://doi.org/10.1103/PhysRevB.85.165120} {\bibfield  {journal} {\bibinfo
  {journal} {Phys. Rev. B}\ }\textbf {\bibinfo {volume} {85}},\ \bibinfo
  {pages} {165120} (\bibinfo {year} {2012})}\BibitemShut {NoStop}%
\bibitem [{\citenamefont {Fang}\ and\ \citenamefont
  {Fu}(2015)}]{PhysRevB.91.161105}%
  \BibitemOpen
  \bibfield  {author} {\bibinfo {author} {\bibfnamefont {C.}~\bibnamefont
  {Fang}}\ and\ \bibinfo {author} {\bibfnamefont {L.}~\bibnamefont {Fu}},\
  }\bibfield  {title} {\bibinfo {title} {New classes of three-dimensional
  topological crystalline insulators: Nonsymmorphic and magnetic},\ }\href
  {https://doi.org/10.1103/PhysRevB.91.161105} {\bibfield  {journal} {\bibinfo
  {journal} {Phys. Rev. B}\ }\textbf {\bibinfo {volume} {91}},\ \bibinfo
  {pages} {161105} (\bibinfo {year} {2015})}\BibitemShut {NoStop}%
\bibitem [{\citenamefont {Ahn}\ \emph {et~al.}(2018)\citenamefont {Ahn},
  \citenamefont {Kim}, \citenamefont {Kim},\ and\ \citenamefont
  {Yang}}]{PhysRevLett.121.106403}%
  \BibitemOpen
  \bibfield  {author} {\bibinfo {author} {\bibfnamefont {J.}~\bibnamefont
  {Ahn}}, \bibinfo {author} {\bibfnamefont {D.}~\bibnamefont {Kim}}, \bibinfo
  {author} {\bibfnamefont {Y.}~\bibnamefont {Kim}},\ and\ \bibinfo {author}
  {\bibfnamefont {B.-J.}\ \bibnamefont {Yang}},\ }\bibfield  {title} {\bibinfo
  {title} {Band topology and linking structure of nodal line semimetals with
  ${Z}_{2}$ monopole charges},\ }\href
  {https://doi.org/10.1103/PhysRevLett.121.106403} {\bibfield  {journal}
  {\bibinfo  {journal} {Phys. Rev. Lett.}\ }\textbf {\bibinfo {volume} {121}},\
  \bibinfo {pages} {106403} (\bibinfo {year} {2018})}\BibitemShut {NoStop}%
\bibitem [{\citenamefont {Ahn}\ and\ \citenamefont
  {Yang}(2019)}]{PhysRevB.99.235125}%
  \BibitemOpen
  \bibfield  {author} {\bibinfo {author} {\bibfnamefont {J.}~\bibnamefont
  {Ahn}}\ and\ \bibinfo {author} {\bibfnamefont {B.-J.}\ \bibnamefont {Yang}},\
  }\bibfield  {title} {\bibinfo {title} {Symmetry representation approach to
  topological invariants in ${C}_{2z}t$-symmetric systems},\ }\href
  {https://doi.org/10.1103/PhysRevB.99.235125} {\bibfield  {journal} {\bibinfo
  {journal} {Phys. Rev. B}\ }\textbf {\bibinfo {volume} {99}},\ \bibinfo
  {pages} {235125} (\bibinfo {year} {2019})}\BibitemShut {NoStop}%
\bibitem [{\citenamefont {Gonz\'alez-Hern\'andez}\ \emph
  {et~al.}(2022)\citenamefont {Gonz\'alez-Hern\'andez}, \citenamefont
  {Pinilla},\ and\ \citenamefont {Uribe}}]{PhysRevB.106.195144}%
  \BibitemOpen
  \bibfield  {author} {\bibinfo {author} {\bibfnamefont {R.}~\bibnamefont
  {Gonz\'alez-Hern\'andez}}, \bibinfo {author} {\bibfnamefont {C.}~\bibnamefont
  {Pinilla}},\ and\ \bibinfo {author} {\bibfnamefont {B.}~\bibnamefont
  {Uribe}},\ }\bibfield  {title} {\bibinfo {title} {Axion insulators protected
  by ${C}_{2}\mathbb{T}$ symmetry, their $k$-theory invariants, and material
  realizations},\ }\href {https://doi.org/10.1103/PhysRevB.106.195144}
  {\bibfield  {journal} {\bibinfo  {journal} {Phys. Rev. B}\ }\textbf {\bibinfo
  {volume} {106}},\ \bibinfo {pages} {195144} (\bibinfo {year}
  {2022})}\BibitemShut {NoStop}%
\bibitem [{\citenamefont {Mong}\ \emph {et~al.}(2010)\citenamefont {Mong},
  \citenamefont {Essin},\ and\ \citenamefont {Moore}}]{PhysRevB.81.245209}%
  \BibitemOpen
  \bibfield  {author} {\bibinfo {author} {\bibfnamefont {R.~S.~K.}\
  \bibnamefont {Mong}}, \bibinfo {author} {\bibfnamefont {A.~M.}\ \bibnamefont
  {Essin}},\ and\ \bibinfo {author} {\bibfnamefont {J.~E.}\ \bibnamefont
  {Moore}},\ }\bibfield  {title} {\bibinfo {title} {Antiferromagnetic
  topological insulators},\ }\href {https://doi.org/10.1103/PhysRevB.81.245209}
  {\bibfield  {journal} {\bibinfo  {journal} {Phys. Rev. B}\ }\textbf {\bibinfo
  {volume} {81}},\ \bibinfo {pages} {245209} (\bibinfo {year}
  {2010})}\BibitemShut {NoStop}%
\bibitem [{\citenamefont {Zhang}\ \emph {et~al.}(2019)\citenamefont {Zhang},
  \citenamefont {Shi}, \citenamefont {Zhu}, \citenamefont {Xing}, \citenamefont
  {Zhang},\ and\ \citenamefont {Wang}}]{PhysRevLett.122.206401}%
  \BibitemOpen
  \bibfield  {author} {\bibinfo {author} {\bibfnamefont {D.}~\bibnamefont
  {Zhang}}, \bibinfo {author} {\bibfnamefont {M.}~\bibnamefont {Shi}}, \bibinfo
  {author} {\bibfnamefont {T.}~\bibnamefont {Zhu}}, \bibinfo {author}
  {\bibfnamefont {D.}~\bibnamefont {Xing}}, \bibinfo {author} {\bibfnamefont
  {H.}~\bibnamefont {Zhang}},\ and\ \bibinfo {author} {\bibfnamefont
  {J.}~\bibnamefont {Wang}},\ }\bibfield  {title} {\bibinfo {title}
  {Topological axion states in the magnetic insulator
  ${\mathrm{mnbi}}_{2}{\mathrm{te}}_{4}$ with the quantized magnetoelectric
  effect},\ }\href {https://doi.org/10.1103/PhysRevLett.122.206401} {\bibfield
  {journal} {\bibinfo  {journal} {Phys. Rev. Lett.}\ }\textbf {\bibinfo
  {volume} {122}},\ \bibinfo {pages} {206401} (\bibinfo {year}
  {2019})}\BibitemShut {NoStop}%
\bibitem [{\citenamefont {Li}\ \emph {et~al.}(2019)\citenamefont {Li},
  \citenamefont {Li}, \citenamefont {Du}, \citenamefont {Wang}, \citenamefont
  {Gu}, \citenamefont {Zhang}, \citenamefont {He}, \citenamefont {Duan},\ and\
  \citenamefont {Xu}}]{li2019intrinsic}%
  \BibitemOpen
  \bibfield  {author} {\bibinfo {author} {\bibfnamefont {J.}~\bibnamefont
  {Li}}, \bibinfo {author} {\bibfnamefont {Y.}~\bibnamefont {Li}}, \bibinfo
  {author} {\bibfnamefont {S.}~\bibnamefont {Du}}, \bibinfo {author}
  {\bibfnamefont {Z.}~\bibnamefont {Wang}}, \bibinfo {author} {\bibfnamefont
  {B.-L.}\ \bibnamefont {Gu}}, \bibinfo {author} {\bibfnamefont {S.-C.}\
  \bibnamefont {Zhang}}, \bibinfo {author} {\bibfnamefont {K.}~\bibnamefont
  {He}}, \bibinfo {author} {\bibfnamefont {W.}~\bibnamefont {Duan}},\ and\
  \bibinfo {author} {\bibfnamefont {Y.}~\bibnamefont {Xu}},\ }\bibfield
  {title} {\bibinfo {title} {Intrinsic magnetic topological insulators in van
  der waals layered mnbi2te4-family materials},\ }\href@noop {} {\bibfield
  {journal} {\bibinfo  {journal} {Science Advances}\ }\textbf {\bibinfo
  {volume} {5}},\ \bibinfo {pages} {eaaw5685} (\bibinfo {year}
  {2019})}\BibitemShut {NoStop}%
\bibitem [{\citenamefont {Qi}\ \emph {et~al.}(2008)\citenamefont {Qi},
  \citenamefont {Hughes},\ and\ \citenamefont {Zhang}}]{PhysRevB.78.195424}%
  \BibitemOpen
  \bibfield  {author} {\bibinfo {author} {\bibfnamefont {X.-L.}\ \bibnamefont
  {Qi}}, \bibinfo {author} {\bibfnamefont {T.~L.}\ \bibnamefont {Hughes}},\
  and\ \bibinfo {author} {\bibfnamefont {S.-C.}\ \bibnamefont {Zhang}},\
  }\bibfield  {title} {\bibinfo {title} {Topological field theory of
  time-reversal invariant insulators},\ }\href
  {https://doi.org/10.1103/PhysRevB.78.195424} {\bibfield  {journal} {\bibinfo
  {journal} {Phys. Rev. B}\ }\textbf {\bibinfo {volume} {78}},\ \bibinfo
  {pages} {195424} (\bibinfo {year} {2008})}\BibitemShut {NoStop}%
\bibitem [{\citenamefont {Essin}\ \emph {et~al.}(2009)\citenamefont {Essin},
  \citenamefont {Moore},\ and\ \citenamefont
  {Vanderbilt}}]{PhysRevLett.102.146805}%
  \BibitemOpen
  \bibfield  {author} {\bibinfo {author} {\bibfnamefont {A.~M.}\ \bibnamefont
  {Essin}}, \bibinfo {author} {\bibfnamefont {J.~E.}\ \bibnamefont {Moore}},\
  and\ \bibinfo {author} {\bibfnamefont {D.}~\bibnamefont {Vanderbilt}},\
  }\bibfield  {title} {\bibinfo {title} {Magnetoelectric polarizability and
  axion electrodynamics in crystalline insulators},\ }\href
  {https://doi.org/10.1103/PhysRevLett.102.146805} {\bibfield  {journal}
  {\bibinfo  {journal} {Phys. Rev. Lett.}\ }\textbf {\bibinfo {volume} {102}},\
  \bibinfo {pages} {146805} (\bibinfo {year} {2009})}\BibitemShut {NoStop}%
\bibitem [{\citenamefont {Armitage}\ and\ \citenamefont
  {Wu}(2019)}]{armitage2019matter}%
  \BibitemOpen
  \bibfield  {author} {\bibinfo {author} {\bibfnamefont {N.~P.}\ \bibnamefont
  {Armitage}}\ and\ \bibinfo {author} {\bibfnamefont {L.}~\bibnamefont {Wu}},\
  }\bibfield  {title} {\bibinfo {title} {On the matter of topological
  insulators as magnetoelectrics},\ }\href@noop {} {\bibfield  {journal}
  {\bibinfo  {journal} {SciPost Physics}\ }\textbf {\bibinfo {volume} {6}},\
  \bibinfo {pages} {046} (\bibinfo {year} {2019})}\BibitemShut {NoStop}%
\bibitem [{\citenamefont {Fang}\ \emph {et~al.}(2012)\citenamefont {Fang},
  \citenamefont {Gilbert},\ and\ \citenamefont {Bernevig}}]{fang2012bulk}%
  \BibitemOpen
  \bibfield  {author} {\bibinfo {author} {\bibfnamefont {C.}~\bibnamefont
  {Fang}}, \bibinfo {author} {\bibfnamefont {M.~J.}\ \bibnamefont {Gilbert}},\
  and\ \bibinfo {author} {\bibfnamefont {B.~A.}\ \bibnamefont {Bernevig}},\
  }\bibfield  {title} {\bibinfo {title} {Bulk topological invariants in
  noninteracting point group symmetric insulators},\ }\href@noop {} {\bibfield
  {journal} {\bibinfo  {journal} {Physical Review B}\ }\textbf {\bibinfo
  {volume} {86}},\ \bibinfo {pages} {115112} (\bibinfo {year}
  {2012})}\BibitemShut {NoStop}%
\bibitem [{\citenamefont {Sekine}\ and\ \citenamefont
  {Nomura}(2021)}]{10.1063/5.0038804}%
  \BibitemOpen
  \bibfield  {author} {\bibinfo {author} {\bibfnamefont {A.}~\bibnamefont
  {Sekine}}\ and\ \bibinfo {author} {\bibfnamefont {K.}~\bibnamefont
  {Nomura}},\ }\bibfield  {title} {\bibinfo {title} {{Axion electrodynamics in
  topological materials}},\ }\href {https://doi.org/10.1063/5.0038804}
  {\bibfield  {journal} {\bibinfo  {journal} {Journal of Applied Physics}\
  }\textbf {\bibinfo {volume} {129}},\ \bibinfo {pages} {141101} (\bibinfo
  {year} {2021})}\BibitemShut {NoStop}%
\bibitem [{\citenamefont {Liu}\ \emph {et~al.}(2020)\citenamefont {Liu},
  \citenamefont {Wang}, \citenamefont {Li}, \citenamefont {Wu}, \citenamefont
  {Li}, \citenamefont {Li}, \citenamefont {He}, \citenamefont {Xu},
  \citenamefont {Zhang},\ and\ \citenamefont {Wang}}]{liu2020robust}%
  \BibitemOpen
  \bibfield  {author} {\bibinfo {author} {\bibfnamefont {C.}~\bibnamefont
  {Liu}}, \bibinfo {author} {\bibfnamefont {Y.}~\bibnamefont {Wang}}, \bibinfo
  {author} {\bibfnamefont {H.}~\bibnamefont {Li}}, \bibinfo {author}
  {\bibfnamefont {Y.}~\bibnamefont {Wu}}, \bibinfo {author} {\bibfnamefont
  {Y.}~\bibnamefont {Li}}, \bibinfo {author} {\bibfnamefont {J.}~\bibnamefont
  {Li}}, \bibinfo {author} {\bibfnamefont {K.}~\bibnamefont {He}}, \bibinfo
  {author} {\bibfnamefont {Y.}~\bibnamefont {Xu}}, \bibinfo {author}
  {\bibfnamefont {J.}~\bibnamefont {Zhang}},\ and\ \bibinfo {author}
  {\bibfnamefont {Y.}~\bibnamefont {Wang}},\ }\bibfield  {title} {\bibinfo
  {title} {Robust axion insulator and chern insulator phases in a
  two-dimensional antiferromagnetic topological insulator},\ }\href@noop {}
  {\bibfield  {journal} {\bibinfo  {journal} {Nature materials}\ }\textbf
  {\bibinfo {volume} {19}},\ \bibinfo {pages} {522} (\bibinfo {year}
  {2020})}\BibitemShut {NoStop}%
\bibitem [{\citenamefont {Varnava}\ and\ \citenamefont
  {Vanderbilt}(2018)}]{PhysRevB.98.245117}%
  \BibitemOpen
  \bibfield  {author} {\bibinfo {author} {\bibfnamefont {N.}~\bibnamefont
  {Varnava}}\ and\ \bibinfo {author} {\bibfnamefont {D.}~\bibnamefont
  {Vanderbilt}},\ }\bibfield  {title} {\bibinfo {title} {Surfaces of axion
  insulators},\ }\href {https://doi.org/10.1103/PhysRevB.98.245117} {\bibfield
  {journal} {\bibinfo  {journal} {Phys. Rev. B}\ }\textbf {\bibinfo {volume}
  {98}},\ \bibinfo {pages} {245117} (\bibinfo {year} {2018})}\BibitemShut
  {NoStop}%
\bibitem [{\citenamefont {Provost}\ and\ \citenamefont
  {Vallee}(1980)}]{provost1980riemannian}%
  \BibitemOpen
  \bibfield  {author} {\bibinfo {author} {\bibfnamefont {J.}~\bibnamefont
  {Provost}}\ and\ \bibinfo {author} {\bibfnamefont {G.}~\bibnamefont
  {Vallee}},\ }\bibfield  {title} {\bibinfo {title} {Riemannian structure on
  manifolds of quantum states},\ }\href@noop {} {\bibfield  {journal} {\bibinfo
   {journal} {Communications in Mathematical Physics}\ }\textbf {\bibinfo
  {volume} {76}},\ \bibinfo {pages} {289} (\bibinfo {year} {1980})}\BibitemShut
  {NoStop}%
\bibitem [{\citenamefont {Xiao}\ \emph {et~al.}(2010)\citenamefont {Xiao},
  \citenamefont {Chang},\ and\ \citenamefont {Niu}}]{RevModPhys.82.1959}%
  \BibitemOpen
  \bibfield  {author} {\bibinfo {author} {\bibfnamefont {D.}~\bibnamefont
  {Xiao}}, \bibinfo {author} {\bibfnamefont {M.-C.}\ \bibnamefont {Chang}},\
  and\ \bibinfo {author} {\bibfnamefont {Q.}~\bibnamefont {Niu}},\ }\bibfield
  {title} {\bibinfo {title} {Berry phase effects on electronic properties},\
  }\href {https://doi.org/10.1103/RevModPhys.82.1959} {\bibfield  {journal}
  {\bibinfo  {journal} {Rev. Mod. Phys.}\ }\textbf {\bibinfo {volume} {82}},\
  \bibinfo {pages} {1959} (\bibinfo {year} {2010})}\BibitemShut {NoStop}%
\bibitem [{\citenamefont {Gu}\ \emph {et~al.}(2021)\citenamefont {Gu},
  \citenamefont {Li}, \citenamefont {Sun}, \citenamefont {Zhao}, \citenamefont
  {Liu}, \citenamefont {Liu}, \citenamefont {Lu},\ and\ \citenamefont
  {Liu}}]{gu2021spectral}%
  \BibitemOpen
  \bibfield  {author} {\bibinfo {author} {\bibfnamefont {M.}~\bibnamefont
  {Gu}}, \bibinfo {author} {\bibfnamefont {J.}~\bibnamefont {Li}}, \bibinfo
  {author} {\bibfnamefont {H.}~\bibnamefont {Sun}}, \bibinfo {author}
  {\bibfnamefont {Y.}~\bibnamefont {Zhao}}, \bibinfo {author} {\bibfnamefont
  {C.}~\bibnamefont {Liu}}, \bibinfo {author} {\bibfnamefont {J.}~\bibnamefont
  {Liu}}, \bibinfo {author} {\bibfnamefont {H.}~\bibnamefont {Lu}},\ and\
  \bibinfo {author} {\bibfnamefont {Q.}~\bibnamefont {Liu}},\ }\bibfield
  {title} {\bibinfo {title} {Spectral signatures of the surface anomalous hall
  effect in magnetic axion insulators},\ }\href@noop {} {\bibfield  {journal}
  {\bibinfo  {journal} {Nature communications}\ }\textbf {\bibinfo {volume}
  {12}},\ \bibinfo {pages} {3524} (\bibinfo {year} {2021})}\BibitemShut
  {NoStop}%
\bibitem [{\citenamefont {Wang}\ \emph {et~al.}(2021)\citenamefont {Wang},
  \citenamefont {Gao},\ and\ \citenamefont {Xiao}}]{PhysRevLett.127.277201}%
  \BibitemOpen
  \bibfield  {author} {\bibinfo {author} {\bibfnamefont {C.}~\bibnamefont
  {Wang}}, \bibinfo {author} {\bibfnamefont {Y.}~\bibnamefont {Gao}},\ and\
  \bibinfo {author} {\bibfnamefont {D.}~\bibnamefont {Xiao}},\ }\bibfield
  {title} {\bibinfo {title} {Intrinsic nonlinear hall effect in
  antiferromagnetic tetragonal cumnas},\ }\href
  {https://doi.org/10.1103/PhysRevLett.127.277201} {\bibfield  {journal}
  {\bibinfo  {journal} {Phys. Rev. Lett.}\ }\textbf {\bibinfo {volume} {127}},\
  \bibinfo {pages} {277201} (\bibinfo {year} {2021})}\BibitemShut {NoStop}%
\bibitem [{\citenamefont {Liu}\ \emph {et~al.}(2021)\citenamefont {Liu},
  \citenamefont {Zhao}, \citenamefont {Huang}, \citenamefont {Wu},
  \citenamefont {Sheng}, \citenamefont {Xiao},\ and\ \citenamefont
  {Yang}}]{PhysRevLett.127.277202}%
  \BibitemOpen
  \bibfield  {author} {\bibinfo {author} {\bibfnamefont {H.}~\bibnamefont
  {Liu}}, \bibinfo {author} {\bibfnamefont {J.}~\bibnamefont {Zhao}}, \bibinfo
  {author} {\bibfnamefont {Y.-X.}\ \bibnamefont {Huang}}, \bibinfo {author}
  {\bibfnamefont {W.}~\bibnamefont {Wu}}, \bibinfo {author} {\bibfnamefont
  {X.-L.}\ \bibnamefont {Sheng}}, \bibinfo {author} {\bibfnamefont
  {C.}~\bibnamefont {Xiao}},\ and\ \bibinfo {author} {\bibfnamefont {S.~A.}\
  \bibnamefont {Yang}},\ }\bibfield  {title} {\bibinfo {title} {Intrinsic
  second-order anomalous hall effect and its application in compensated
  antiferromagnets},\ }\href {https://doi.org/10.1103/PhysRevLett.127.277202}
  {\bibfield  {journal} {\bibinfo  {journal} {Phys. Rev. Lett.}\ }\textbf
  {\bibinfo {volume} {127}},\ \bibinfo {pages} {277202} (\bibinfo {year}
  {2021})}\BibitemShut {NoStop}%
\bibitem [{\citenamefont {Wang}\ \emph {et~al.}(2023)\citenamefont {Wang},
  \citenamefont {Kaplan}, \citenamefont {Zhang}, \citenamefont {Holder},
  \citenamefont {Cao}, \citenamefont {Wang}, \citenamefont {Zhou},
  \citenamefont {Zhou}, \citenamefont {Jiang}, \citenamefont {Zhang} \emph
  {et~al.}}]{wang2023quantum}%
  \BibitemOpen
  \bibfield  {author} {\bibinfo {author} {\bibfnamefont {N.}~\bibnamefont
  {Wang}}, \bibinfo {author} {\bibfnamefont {D.}~\bibnamefont {Kaplan}},
  \bibinfo {author} {\bibfnamefont {Z.}~\bibnamefont {Zhang}}, \bibinfo
  {author} {\bibfnamefont {T.}~\bibnamefont {Holder}}, \bibinfo {author}
  {\bibfnamefont {N.}~\bibnamefont {Cao}}, \bibinfo {author} {\bibfnamefont
  {A.}~\bibnamefont {Wang}}, \bibinfo {author} {\bibfnamefont {X.}~\bibnamefont
  {Zhou}}, \bibinfo {author} {\bibfnamefont {F.}~\bibnamefont {Zhou}}, \bibinfo
  {author} {\bibfnamefont {Z.}~\bibnamefont {Jiang}}, \bibinfo {author}
  {\bibfnamefont {C.}~\bibnamefont {Zhang}}, \emph {et~al.},\ }\bibfield
  {title} {\bibinfo {title} {Quantum-metric-induced nonlinear transport in a
  topological antiferromagnet},\ }\href@noop {} {\bibfield  {journal} {\bibinfo
   {journal} {Nature}\ }\textbf {\bibinfo {volume} {621}},\ \bibinfo {pages}
  {487} (\bibinfo {year} {2023})}\BibitemShut {NoStop}%
\bibitem [{\citenamefont {Gao}\ \emph {et~al.}(2023)\citenamefont {Gao},
  \citenamefont {Liu}, \citenamefont {Qiu}, \citenamefont {Ghosh},
  \citenamefont {V.~Trevisan}, \citenamefont {Onishi}, \citenamefont {Hu},
  \citenamefont {Qian}, \citenamefont {Tien}, \citenamefont {Chen} \emph
  {et~al.}}]{gao2023quantum}%
  \BibitemOpen
  \bibfield  {author} {\bibinfo {author} {\bibfnamefont {A.}~\bibnamefont
  {Gao}}, \bibinfo {author} {\bibfnamefont {Y.-F.}\ \bibnamefont {Liu}},
  \bibinfo {author} {\bibfnamefont {J.-X.}\ \bibnamefont {Qiu}}, \bibinfo
  {author} {\bibfnamefont {B.}~\bibnamefont {Ghosh}}, \bibinfo {author}
  {\bibfnamefont {T.}~\bibnamefont {V.~Trevisan}}, \bibinfo {author}
  {\bibfnamefont {Y.}~\bibnamefont {Onishi}}, \bibinfo {author} {\bibfnamefont
  {C.}~\bibnamefont {Hu}}, \bibinfo {author} {\bibfnamefont {T.}~\bibnamefont
  {Qian}}, \bibinfo {author} {\bibfnamefont {H.-J.}\ \bibnamefont {Tien}},
  \bibinfo {author} {\bibfnamefont {S.-W.}\ \bibnamefont {Chen}}, \emph
  {et~al.},\ }\bibfield  {title} {\bibinfo {title} {Quantum metric nonlinear
  hall effect in a topological antiferromagnetic heterostructure},\ }\href@noop
  {} {\bibfield  {journal} {\bibinfo  {journal} {Science}\ }\textbf {\bibinfo
  {volume} {381}},\ \bibinfo {pages} {181} (\bibinfo {year}
  {2023})}\BibitemShut {NoStop}%
\bibitem [{\citenamefont {Kaplan}\ \emph {et~al.}(2024)\citenamefont {Kaplan},
  \citenamefont {Holder},\ and\ \citenamefont {Yan}}]{PhysRevLett.132.026301}%
  \BibitemOpen
  \bibfield  {author} {\bibinfo {author} {\bibfnamefont {D.}~\bibnamefont
  {Kaplan}}, \bibinfo {author} {\bibfnamefont {T.}~\bibnamefont {Holder}},\
  and\ \bibinfo {author} {\bibfnamefont {B.}~\bibnamefont {Yan}},\ }\bibfield
  {title} {\bibinfo {title} {Unification of nonlinear anomalous hall effect and
  nonreciprocal magnetoresistance in metals by the quantum geometry},\ }\href
  {https://doi.org/10.1103/PhysRevLett.132.026301} {\bibfield  {journal}
  {\bibinfo  {journal} {Phys. Rev. Lett.}\ }\textbf {\bibinfo {volume} {132}},\
  \bibinfo {pages} {026301} (\bibinfo {year} {2024})}\BibitemShut {NoStop}%
\bibitem [{\citenamefont {Childs}\ \emph {et~al.}(2019)\citenamefont {Childs},
  \citenamefont {Baranets},\ and\ \citenamefont {Bobev}}]{CHILDS2019120889}%
  \BibitemOpen
  \bibfield  {author} {\bibinfo {author} {\bibfnamefont {A.~B.}\ \bibnamefont
  {Childs}}, \bibinfo {author} {\bibfnamefont {S.}~\bibnamefont {Baranets}},\
  and\ \bibinfo {author} {\bibfnamefont {S.}~\bibnamefont {Bobev}},\ }\bibfield
   {title} {\bibinfo {title} {Five new ternary indium-arsenides discovered.
  synthesis and structural characterization of the zintl phases sr3in2as4,
  ba3in2as4, eu3in2as4, sr5in2as6 and eu5in2as6},\ }\href
  {https://doi.org/https://doi.org/10.1016/j.jssc.2019.07.050} {\bibfield
  {journal} {\bibinfo  {journal} {Journal of Solid State Chemistry}\ }\textbf
  {\bibinfo {volume} {278}},\ \bibinfo {pages} {120889} (\bibinfo {year}
  {2019})}\BibitemShut {NoStop}%
\bibitem [{\citenamefont {Zhang}\ \emph
  {et~al.}(2020{\natexlab{b}})\citenamefont {Zhang}, \citenamefont {Deng},
  \citenamefont {Zhang}, \citenamefont {Wang}, \citenamefont {Wang},
  \citenamefont {Liu}, \citenamefont {Mei}, \citenamefont {Kumar},
  \citenamefont {Schwier}, \citenamefont {Shimada}, \citenamefont {Chen},\ and\
  \citenamefont {Shen}}]{PhysRevB.101.205126}%
  \BibitemOpen
  \bibfield  {author} {\bibinfo {author} {\bibfnamefont {Y.}~\bibnamefont
  {Zhang}}, \bibinfo {author} {\bibfnamefont {K.}~\bibnamefont {Deng}},
  \bibinfo {author} {\bibfnamefont {X.}~\bibnamefont {Zhang}}, \bibinfo
  {author} {\bibfnamefont {M.}~\bibnamefont {Wang}}, \bibinfo {author}
  {\bibfnamefont {Y.}~\bibnamefont {Wang}}, \bibinfo {author} {\bibfnamefont
  {C.}~\bibnamefont {Liu}}, \bibinfo {author} {\bibfnamefont {J.-W.}\
  \bibnamefont {Mei}}, \bibinfo {author} {\bibfnamefont {S.}~\bibnamefont
  {Kumar}}, \bibinfo {author} {\bibfnamefont {E.~F.}\ \bibnamefont {Schwier}},
  \bibinfo {author} {\bibfnamefont {K.}~\bibnamefont {Shimada}}, \bibinfo
  {author} {\bibfnamefont {C.}~\bibnamefont {Chen}},\ and\ \bibinfo {author}
  {\bibfnamefont {B.}~\bibnamefont {Shen}},\ }\bibfield  {title} {\bibinfo
  {title} {In-plane antiferromagnetic moments and magnetic polaron in the axion
  topological insulator candidate ${\mathrm{euin}}_{2}{\mathrm{as}}_{2}$},\
  }\href {https://doi.org/10.1103/PhysRevB.101.205126} {\bibfield  {journal}
  {\bibinfo  {journal} {Phys. Rev. B}\ }\textbf {\bibinfo {volume} {101}},\
  \bibinfo {pages} {205126} (\bibinfo {year} {2020}{\natexlab{b}})}\BibitemShut
  {NoStop}%
\bibitem [{\citenamefont {Sato}\ \emph {et~al.}(2020)\citenamefont {Sato},
  \citenamefont {Wang}, \citenamefont {Takane}, \citenamefont {Souma},
  \citenamefont {Cui}, \citenamefont {Li}, \citenamefont {Nakayama},
  \citenamefont {Kawakami}, \citenamefont {Kubota}, \citenamefont {Cacho},
  \citenamefont {Kim}, \citenamefont {Arab}, \citenamefont {Strocov},
  \citenamefont {Yao},\ and\ \citenamefont
  {Takahashi}}]{PhysRevResearch.2.033342}%
  \BibitemOpen
  \bibfield  {author} {\bibinfo {author} {\bibfnamefont {T.}~\bibnamefont
  {Sato}}, \bibinfo {author} {\bibfnamefont {Z.}~\bibnamefont {Wang}}, \bibinfo
  {author} {\bibfnamefont {D.}~\bibnamefont {Takane}}, \bibinfo {author}
  {\bibfnamefont {S.}~\bibnamefont {Souma}}, \bibinfo {author} {\bibfnamefont
  {C.}~\bibnamefont {Cui}}, \bibinfo {author} {\bibfnamefont {Y.}~\bibnamefont
  {Li}}, \bibinfo {author} {\bibfnamefont {K.}~\bibnamefont {Nakayama}},
  \bibinfo {author} {\bibfnamefont {T.}~\bibnamefont {Kawakami}}, \bibinfo
  {author} {\bibfnamefont {Y.}~\bibnamefont {Kubota}}, \bibinfo {author}
  {\bibfnamefont {C.}~\bibnamefont {Cacho}}, \bibinfo {author} {\bibfnamefont
  {T.~K.}\ \bibnamefont {Kim}}, \bibinfo {author} {\bibfnamefont
  {A.}~\bibnamefont {Arab}}, \bibinfo {author} {\bibfnamefont {V.~N.}\
  \bibnamefont {Strocov}}, \bibinfo {author} {\bibfnamefont {Y.}~\bibnamefont
  {Yao}},\ and\ \bibinfo {author} {\bibfnamefont {T.}~\bibnamefont
  {Takahashi}},\ }\bibfield  {title} {\bibinfo {title} {Signature of band
  inversion in the antiferromagnetic phase of axion insulator candidate
  ${\mathrm{euin}}_{2}{\mathrm{as}}_{2}$},\ }\href
  {https://doi.org/10.1103/PhysRevResearch.2.033342} {\bibfield  {journal}
  {\bibinfo  {journal} {Phys. Rev. Res.}\ }\textbf {\bibinfo {volume} {2}},\
  \bibinfo {pages} {033342} (\bibinfo {year} {2020})}\BibitemShut {NoStop}%
\bibitem [{\citenamefont {Riberolles}\ \emph {et~al.}(2021)\citenamefont
  {Riberolles}, \citenamefont {Trevisan}, \citenamefont {Kuthanazhi},
  \citenamefont {Heitmann}, \citenamefont {Ye}, \citenamefont {Johnston},
  \citenamefont {Bud’ko}, \citenamefont {Ryan}, \citenamefont {Canfield},
  \citenamefont {Kreyssig} \emph {et~al.}}]{riberolles2021magnetic}%
  \BibitemOpen
  \bibfield  {author} {\bibinfo {author} {\bibfnamefont {S.~X.}\ \bibnamefont
  {Riberolles}}, \bibinfo {author} {\bibfnamefont {T.~V.}\ \bibnamefont
  {Trevisan}}, \bibinfo {author} {\bibfnamefont {B.}~\bibnamefont
  {Kuthanazhi}}, \bibinfo {author} {\bibfnamefont {T.}~\bibnamefont
  {Heitmann}}, \bibinfo {author} {\bibfnamefont {F.}~\bibnamefont {Ye}},
  \bibinfo {author} {\bibfnamefont {D.}~\bibnamefont {Johnston}}, \bibinfo
  {author} {\bibfnamefont {S.}~\bibnamefont {Bud’ko}}, \bibinfo {author}
  {\bibfnamefont {D.}~\bibnamefont {Ryan}}, \bibinfo {author} {\bibfnamefont
  {P.}~\bibnamefont {Canfield}}, \bibinfo {author} {\bibfnamefont
  {A.}~\bibnamefont {Kreyssig}}, \emph {et~al.},\ }\bibfield  {title} {\bibinfo
  {title} {Magnetic crystalline-symmetry-protected axion electrodynamics and
  field-tunable unpinned dirac cones in euin2as2},\ }\href@noop {} {\bibfield
  {journal} {\bibinfo  {journal} {Nature communications}\ }\textbf {\bibinfo
  {volume} {12}},\ \bibinfo {pages} {999} (\bibinfo {year} {2021})}\BibitemShut
  {NoStop}%
\bibitem [{\citenamefont {Soh}\ \emph {et~al.}(2023)\citenamefont {Soh},
  \citenamefont {Bombardi}, \citenamefont {Mila}, \citenamefont {Rahn},
  \citenamefont {Prabhakaran}, \citenamefont {Francoual}, \citenamefont
  {R{\o}nnow},\ and\ \citenamefont {Boothroyd}}]{soh2023understanding}%
  \BibitemOpen
  \bibfield  {author} {\bibinfo {author} {\bibfnamefont {J.-R.}\ \bibnamefont
  {Soh}}, \bibinfo {author} {\bibfnamefont {A.}~\bibnamefont {Bombardi}},
  \bibinfo {author} {\bibfnamefont {F.}~\bibnamefont {Mila}}, \bibinfo {author}
  {\bibfnamefont {M.~C.}\ \bibnamefont {Rahn}}, \bibinfo {author}
  {\bibfnamefont {D.}~\bibnamefont {Prabhakaran}}, \bibinfo {author}
  {\bibfnamefont {S.}~\bibnamefont {Francoual}}, \bibinfo {author}
  {\bibfnamefont {H.~M.}\ \bibnamefont {R{\o}nnow}},\ and\ \bibinfo {author}
  {\bibfnamefont {A.~T.}\ \bibnamefont {Boothroyd}},\ }\bibfield  {title}
  {\bibinfo {title} {Understanding unconventional magnetic order in a candidate
  axion insulator by resonant elastic x-ray scattering},\ }\href@noop {}
  {\bibfield  {journal} {\bibinfo  {journal} {Nature Communications}\ }\textbf
  {\bibinfo {volume} {14}},\ \bibinfo {pages} {3387} (\bibinfo {year}
  {2023})}\BibitemShut {NoStop}%
\bibitem [{\citenamefont {Donoway}\ \emph {et~al.}(2023)\citenamefont
  {Donoway}, \citenamefont {Trevisan}, \citenamefont {Peláez}, \citenamefont
  {Day}, \citenamefont {Yamakawa}, \citenamefont {Sun}, \citenamefont {Soh},
  \citenamefont {Prabhakaran}, \citenamefont {Boothroyd}, \citenamefont
  {Fernandes}, \citenamefont {Analytis}, \citenamefont {Moore}, \citenamefont
  {Orenstein},\ and\ \citenamefont {Sunko}}]{donoway2023symmetrybreaking}%
  \BibitemOpen
  \bibfield  {author} {\bibinfo {author} {\bibfnamefont {E.}~\bibnamefont
  {Donoway}}, \bibinfo {author} {\bibfnamefont {T.~V.}\ \bibnamefont
  {Trevisan}}, \bibinfo {author} {\bibfnamefont {A.~L.}\ \bibnamefont
  {Peláez}}, \bibinfo {author} {\bibfnamefont {R.~P.}\ \bibnamefont {Day}},
  \bibinfo {author} {\bibfnamefont {K.}~\bibnamefont {Yamakawa}}, \bibinfo
  {author} {\bibfnamefont {Y.}~\bibnamefont {Sun}}, \bibinfo {author}
  {\bibfnamefont {J.~R.}\ \bibnamefont {Soh}}, \bibinfo {author} {\bibfnamefont
  {D.}~\bibnamefont {Prabhakaran}}, \bibinfo {author} {\bibfnamefont {A.~T.}\
  \bibnamefont {Boothroyd}}, \bibinfo {author} {\bibfnamefont {R.~M.}\
  \bibnamefont {Fernandes}}, \bibinfo {author} {\bibfnamefont {J.~G.}\
  \bibnamefont {Analytis}}, \bibinfo {author} {\bibfnamefont {J.~E.}\
  \bibnamefont {Moore}}, \bibinfo {author} {\bibfnamefont {J.}~\bibnamefont
  {Orenstein}},\ and\ \bibinfo {author} {\bibfnamefont {V.}~\bibnamefont
  {Sunko}},\ }\href@noop {} {\bibinfo {title} {Symmetry-breaking pathway
  towards the unpinned broken helix}} (\bibinfo {year} {2023}),\ \Eprint
  {https://arxiv.org/abs/2310.16018} {arXiv:2310.16018 [cond-mat.str-el]}
  \BibitemShut {NoStop}%
\bibitem [{\citenamefont {Rosa}\ \emph {et~al.}(2020)\citenamefont {Rosa},
  \citenamefont {Xu}, \citenamefont {Rahn}, \citenamefont {Souza},
  \citenamefont {Kushwaha}, \citenamefont {Veiga}, \citenamefont {Bombardi},
  \citenamefont {Thomas}, \citenamefont {Janoschek}, \citenamefont {Bauer}
  \emph {et~al.}}]{rosa2020colossal}%
  \BibitemOpen
  \bibfield  {author} {\bibinfo {author} {\bibfnamefont {P.}~\bibnamefont
  {Rosa}}, \bibinfo {author} {\bibfnamefont {Y.}~\bibnamefont {Xu}}, \bibinfo
  {author} {\bibfnamefont {M.}~\bibnamefont {Rahn}}, \bibinfo {author}
  {\bibfnamefont {J.}~\bibnamefont {Souza}}, \bibinfo {author} {\bibfnamefont
  {S.}~\bibnamefont {Kushwaha}}, \bibinfo {author} {\bibfnamefont
  {L.}~\bibnamefont {Veiga}}, \bibinfo {author} {\bibfnamefont
  {A.}~\bibnamefont {Bombardi}}, \bibinfo {author} {\bibfnamefont
  {S.}~\bibnamefont {Thomas}}, \bibinfo {author} {\bibfnamefont
  {M.}~\bibnamefont {Janoschek}}, \bibinfo {author} {\bibfnamefont
  {E.}~\bibnamefont {Bauer}}, \emph {et~al.},\ }\bibfield  {title} {\bibinfo
  {title} {Colossal magnetoresistance in a nonsymmorphic antiferromagnetic
  insulator},\ }\href@noop {} {\bibfield  {journal} {\bibinfo  {journal} {npj
  Quantum Materials}\ }\textbf {\bibinfo {volume} {5}},\ \bibinfo {pages} {52}
  (\bibinfo {year} {2020})}\BibitemShut {NoStop}%
\bibitem [{\citenamefont {Crivillero}\ \emph {et~al.}(2022)\citenamefont
  {Crivillero}, \citenamefont {R{\"o}{\ss}ler}, \citenamefont {Rosa},
  \citenamefont {M{\"u}ller}, \citenamefont {R{\"o}{\ss}ler},\ and\
  \citenamefont {Wirth}}]{crivillero2022surface}%
  \BibitemOpen
  \bibfield  {author} {\bibinfo {author} {\bibfnamefont {M.~V.~A.}\
  \bibnamefont {Crivillero}}, \bibinfo {author} {\bibfnamefont
  {S.}~\bibnamefont {R{\"o}{\ss}ler}}, \bibinfo {author} {\bibfnamefont
  {P.~F.}\ \bibnamefont {Rosa}}, \bibinfo {author} {\bibfnamefont
  {J.}~\bibnamefont {M{\"u}ller}}, \bibinfo {author} {\bibfnamefont
  {U.}~\bibnamefont {R{\"o}{\ss}ler}},\ and\ \bibinfo {author} {\bibfnamefont
  {S.}~\bibnamefont {Wirth}},\ }\bibfield  {title} {\bibinfo {title} {Surface
  and electronic structure at atomic length scales of the nonsymmorphic
  antiferromagnet eu$_{5}$in$_{2}$sb$_{6}$},\ }\href@noop {} {\bibfield
  {journal} {\bibinfo  {journal} {Physical Review B}\ }\textbf {\bibinfo
  {volume} {106}},\ \bibinfo {pages} {035124} (\bibinfo {year}
  {2022})}\BibitemShut {NoStop}%
\bibitem [{\citenamefont {Rahn}\ \emph {et~al.}(2024)\citenamefont {Rahn},
  \citenamefont {Wilson}, \citenamefont {Hicken}, \citenamefont {Pratt},
  \citenamefont {Wang}, \citenamefont {Orlandi}, \citenamefont {Khalyavin},
  \citenamefont {Manuel}, \citenamefont {Veiga}, \citenamefont {Bombardi},
  \citenamefont {Francoual}, \citenamefont {Bereciartua}, \citenamefont
  {Sukhanov}, \citenamefont {Thompson}, \citenamefont {Thomas}, \citenamefont
  {Rosa}, \citenamefont {Lancaster}, \citenamefont {Ronning},\ and\
  \citenamefont {Janoschek}}]{rahn2023unusual}%
  \BibitemOpen
  \bibfield  {author} {\bibinfo {author} {\bibfnamefont {M.~C.}\ \bibnamefont
  {Rahn}}, \bibinfo {author} {\bibfnamefont {M.~N.}\ \bibnamefont {Wilson}},
  \bibinfo {author} {\bibfnamefont {T.~J.}\ \bibnamefont {Hicken}}, \bibinfo
  {author} {\bibfnamefont {F.~L.}\ \bibnamefont {Pratt}}, \bibinfo {author}
  {\bibfnamefont {C.}~\bibnamefont {Wang}}, \bibinfo {author} {\bibfnamefont
  {F.}~\bibnamefont {Orlandi}}, \bibinfo {author} {\bibfnamefont {D.~D.}\
  \bibnamefont {Khalyavin}}, \bibinfo {author} {\bibfnamefont {P.}~\bibnamefont
  {Manuel}}, \bibinfo {author} {\bibfnamefont {L.~S.~I.}\ \bibnamefont
  {Veiga}}, \bibinfo {author} {\bibfnamefont {A.}~\bibnamefont {Bombardi}},
  \bibinfo {author} {\bibfnamefont {S.}~\bibnamefont {Francoual}}, \bibinfo
  {author} {\bibfnamefont {P.}~\bibnamefont {Bereciartua}}, \bibinfo {author}
  {\bibfnamefont {A.~S.}\ \bibnamefont {Sukhanov}}, \bibinfo {author}
  {\bibfnamefont {J.~D.}\ \bibnamefont {Thompson}}, \bibinfo {author}
  {\bibfnamefont {S.~M.}\ \bibnamefont {Thomas}}, \bibinfo {author}
  {\bibfnamefont {P.~F.~S.}\ \bibnamefont {Rosa}}, \bibinfo {author}
  {\bibfnamefont {T.}~\bibnamefont {Lancaster}}, \bibinfo {author}
  {\bibfnamefont {F.}~\bibnamefont {Ronning}},\ and\ \bibinfo {author}
  {\bibfnamefont {M.}~\bibnamefont {Janoschek}},\ }\bibfield  {title} {\bibinfo
  {title} {Magnetism in the axion insulator candidate
  ${\mathrm{eu}}_{5}{\mathrm{in}}_{2}{\mathrm{sb}}_{6}$},\ }\href
  {https://doi.org/10.1103/PhysRevB.109.174404} {\bibfield  {journal} {\bibinfo
   {journal} {Phys. Rev. B}\ }\textbf {\bibinfo {volume} {109}},\ \bibinfo
  {pages} {174404} (\bibinfo {year} {2024})}\BibitemShut {NoStop}%
\bibitem [{\citenamefont {Morano}\ \emph {et~al.}(2024)\citenamefont {Morano},
  \citenamefont {Gaudet}, \citenamefont {Varnava}, \citenamefont {Berry},
  \citenamefont {Halloran}, \citenamefont {Lygouras}, \citenamefont {Wang},
  \citenamefont {Hoffman}, \citenamefont {Xu}, \citenamefont {Lynn},
  \citenamefont {McQueen}, \citenamefont {Vanderbilt},\ and\ \citenamefont
  {Broholm}}]{PhysRevB.109.014432}%
  \BibitemOpen
  \bibfield  {author} {\bibinfo {author} {\bibfnamefont {V.~C.}\ \bibnamefont
  {Morano}}, \bibinfo {author} {\bibfnamefont {J.}~\bibnamefont {Gaudet}},
  \bibinfo {author} {\bibfnamefont {N.}~\bibnamefont {Varnava}}, \bibinfo
  {author} {\bibfnamefont {T.}~\bibnamefont {Berry}}, \bibinfo {author}
  {\bibfnamefont {T.}~\bibnamefont {Halloran}}, \bibinfo {author}
  {\bibfnamefont {C.~J.}\ \bibnamefont {Lygouras}}, \bibinfo {author}
  {\bibfnamefont {X.}~\bibnamefont {Wang}}, \bibinfo {author} {\bibfnamefont
  {C.~M.}\ \bibnamefont {Hoffman}}, \bibinfo {author} {\bibfnamefont
  {G.}~\bibnamefont {Xu}}, \bibinfo {author} {\bibfnamefont {J.~W.}\
  \bibnamefont {Lynn}}, \bibinfo {author} {\bibfnamefont {T.~M.}\ \bibnamefont
  {McQueen}}, \bibinfo {author} {\bibfnamefont {D.}~\bibnamefont
  {Vanderbilt}},\ and\ \bibinfo {author} {\bibfnamefont {C.~L.}\ \bibnamefont
  {Broholm}},\ }\bibfield  {title} {\bibinfo {title} {Noncollinear
  $2\mathrm{k}$ antiferromagnetism in the zintl semiconductor
  ${\mathrm{eu}}_{5}{\mathrm{in}}_{2}{\mathrm{sb}}_{6}$},\ }\href
  {https://doi.org/10.1103/PhysRevB.109.014432} {\bibfield  {journal} {\bibinfo
   {journal} {Phys. Rev. B}\ }\textbf {\bibinfo {volume} {109}},\ \bibinfo
  {pages} {014432} (\bibinfo {year} {2024})}\BibitemShut {NoStop}%
\bibitem [{\citenamefont {{Song}}\ \emph {et~al.}(2024)\citenamefont {{Song}},
  \citenamefont {{Houben}}, \citenamefont {{Zhao}}, \citenamefont {{Bae}},
  \citenamefont {{Rothem}}, \citenamefont {{Gupta}}, \citenamefont {{Yan}},
  \citenamefont {{Kalisky}}, \citenamefont {{Zaluska-Kotur}}, \citenamefont
  {{Kacman}}, \citenamefont {{Shtrikman}},\ and\ \citenamefont
  {{Beidenkopf}}}]{haim324}%
  \BibitemOpen
  \bibfield  {author} {\bibinfo {author} {\bibfnamefont {M.~S.}\ \bibnamefont
  {{Song}}}, \bibinfo {author} {\bibfnamefont {L.}~\bibnamefont {{Houben}}},
  \bibinfo {author} {\bibfnamefont {Y.}~\bibnamefont {{Zhao}}}, \bibinfo
  {author} {\bibfnamefont {H.}~\bibnamefont {{Bae}}}, \bibinfo {author}
  {\bibfnamefont {N.}~\bibnamefont {{Rothem}}}, \bibinfo {author}
  {\bibfnamefont {A.}~\bibnamefont {{Gupta}}}, \bibinfo {author} {\bibfnamefont
  {B.}~\bibnamefont {{Yan}}}, \bibinfo {author} {\bibfnamefont
  {B.}~\bibnamefont {{Kalisky}}}, \bibinfo {author} {\bibfnamefont
  {M.}~\bibnamefont {{Zaluska-Kotur}}}, \bibinfo {author} {\bibfnamefont
  {P.}~\bibnamefont {{Kacman}}}, \bibinfo {author} {\bibfnamefont
  {H.}~\bibnamefont {{Shtrikman}}},\ and\ \bibinfo {author} {\bibfnamefont
  {H.}~\bibnamefont {{Beidenkopf}}},\ }\bibfield  {title} {\bibinfo {title}
  {{Topotaxial Mutual-Exchange Growth of Magnetic Zintl Eu$_3$In$_2$As$_4$
  Nanowires with Axion Insulator Classification}},\ }\href@noop {} {\bibfield
  {journal} {\bibinfo  {journal} {arXiv e-prints}\ } (\bibinfo {year}
  {2024})},\ \Eprint {https://arxiv.org/abs/2406.18956} {arXiv:2406.18956
  [cond-mat.mtrl-sci]} \BibitemShut {NoStop}%
\bibitem [{\citenamefont {Jia}\ \emph {et~al.}(2024)\citenamefont {Jia},
  \citenamefont {Yao}, \citenamefont {He}, \citenamefont {Li}, \citenamefont
  {Deng}, \citenamefont {Yang}, \citenamefont {Wang}, \citenamefont {Zhu},
  \citenamefont {Wang}, \citenamefont {Yan}, \citenamefont {Feng},
  \citenamefont {Shen}, \citenamefont {Luo}, \citenamefont {Wang},\ and\
  \citenamefont {Shi}}]{jia2024discovery}%
  \BibitemOpen
  \bibfield  {author} {\bibinfo {author} {\bibfnamefont {K.}~\bibnamefont
  {Jia}}, \bibinfo {author} {\bibfnamefont {J.}~\bibnamefont {Yao}}, \bibinfo
  {author} {\bibfnamefont {X.}~\bibnamefont {He}}, \bibinfo {author}
  {\bibfnamefont {Y.}~\bibnamefont {Li}}, \bibinfo {author} {\bibfnamefont
  {J.}~\bibnamefont {Deng}}, \bibinfo {author} {\bibfnamefont {M.}~\bibnamefont
  {Yang}}, \bibinfo {author} {\bibfnamefont {J.}~\bibnamefont {Wang}}, \bibinfo
  {author} {\bibfnamefont {Z.}~\bibnamefont {Zhu}}, \bibinfo {author}
  {\bibfnamefont {C.}~\bibnamefont {Wang}}, \bibinfo {author} {\bibfnamefont
  {D.}~\bibnamefont {Yan}}, \bibinfo {author} {\bibfnamefont {H.~L.}\
  \bibnamefont {Feng}}, \bibinfo {author} {\bibfnamefont {J.}~\bibnamefont
  {Shen}}, \bibinfo {author} {\bibfnamefont {Y.}~\bibnamefont {Luo}}, \bibinfo
  {author} {\bibfnamefont {Z.}~\bibnamefont {Wang}},\ and\ \bibinfo {author}
  {\bibfnamefont {Y.}~\bibnamefont {Shi}},\ }\href@noop {} {\bibinfo {title}
  {Discovery of a magnetic topological semimetal eu$_3$in$_2$as$_4$ with a
  single pair of weyl points}} (\bibinfo {year} {2024}),\ \Eprint
  {https://arxiv.org/abs/2403.07637} {arXiv:2403.07637 [cond-mat.mes-hall]}
  \BibitemShut {NoStop}%
\bibitem [{\citenamefont {Liu}\ \emph {et~al.}(2022)\citenamefont {Liu},
  \citenamefont {Li}, \citenamefont {Han}, \citenamefont {Wan},\ and\
  \citenamefont {Liu}}]{PhysRevX.12.021016}%
  \BibitemOpen
  \bibfield  {author} {\bibinfo {author} {\bibfnamefont {P.}~\bibnamefont
  {Liu}}, \bibinfo {author} {\bibfnamefont {J.}~\bibnamefont {Li}}, \bibinfo
  {author} {\bibfnamefont {J.}~\bibnamefont {Han}}, \bibinfo {author}
  {\bibfnamefont {X.}~\bibnamefont {Wan}},\ and\ \bibinfo {author}
  {\bibfnamefont {Q.}~\bibnamefont {Liu}},\ }\bibfield  {title} {\bibinfo
  {title} {Spin-group symmetry in magnetic materials with negligible spin-orbit
  coupling},\ }\href {https://doi.org/10.1103/PhysRevX.12.021016} {\bibfield
  {journal} {\bibinfo  {journal} {Phys. Rev. X}\ }\textbf {\bibinfo {volume}
  {12}},\ \bibinfo {pages} {021016} (\bibinfo {year} {2022})}\BibitemShut
  {NoStop}%
\bibitem [{\citenamefont {\ifmmode~\check{S}\else \v{S}\fi{}mejkal}\ \emph
  {et~al.}(2022{\natexlab{a}})\citenamefont {\ifmmode~\check{S}\else
  \v{S}\fi{}mejkal}, \citenamefont {Sinova},\ and\ \citenamefont
  {Jungwirth}}]{PhysRevX.12.031042}%
  \BibitemOpen
  \bibfield  {author} {\bibinfo {author} {\bibfnamefont {L.}~\bibnamefont
  {\ifmmode~\check{S}\else \v{S}\fi{}mejkal}}, \bibinfo {author} {\bibfnamefont
  {J.}~\bibnamefont {Sinova}},\ and\ \bibinfo {author} {\bibfnamefont
  {T.}~\bibnamefont {Jungwirth}},\ }\bibfield  {title} {\bibinfo {title}
  {Beyond conventional ferromagnetism and antiferromagnetism: A phase with
  nonrelativistic spin and crystal rotation symmetry},\ }\href
  {https://doi.org/10.1103/PhysRevX.12.031042} {\bibfield  {journal} {\bibinfo
  {journal} {Phys. Rev. X}\ }\textbf {\bibinfo {volume} {12}},\ \bibinfo
  {pages} {031042} (\bibinfo {year} {2022}{\natexlab{a}})}\BibitemShut
  {NoStop}%
\bibitem [{\citenamefont {\ifmmode~\check{S}\else \v{S}\fi{}mejkal}\ \emph
  {et~al.}(2022{\natexlab{b}})\citenamefont {\ifmmode~\check{S}\else
  \v{S}\fi{}mejkal}, \citenamefont {Sinova},\ and\ \citenamefont
  {Jungwirth}}]{PhysRevX.12.040501}%
  \BibitemOpen
  \bibfield  {author} {\bibinfo {author} {\bibfnamefont {L.}~\bibnamefont
  {\ifmmode~\check{S}\else \v{S}\fi{}mejkal}}, \bibinfo {author} {\bibfnamefont
  {J.}~\bibnamefont {Sinova}},\ and\ \bibinfo {author} {\bibfnamefont
  {T.}~\bibnamefont {Jungwirth}},\ }\bibfield  {title} {\bibinfo {title}
  {Emerging research landscape of altermagnetism},\ }\href
  {https://doi.org/10.1103/PhysRevX.12.040501} {\bibfield  {journal} {\bibinfo
  {journal} {Phys. Rev. X}\ }\textbf {\bibinfo {volume} {12}},\ \bibinfo
  {pages} {040501} (\bibinfo {year} {2022}{\natexlab{b}})}\BibitemShut
  {NoStop}%
\bibitem [{\citenamefont {Krempask{\`y}}\ \emph {et~al.}(2024)\citenamefont
  {Krempask{\`y}}, \citenamefont {{\v{S}}mejkal}, \citenamefont {D’Souza},
  \citenamefont {Hajlaoui}, \citenamefont {Springholz}, \citenamefont
  {Uhl{\'\i}{\v{r}}ov{\'a}}, \citenamefont {Alarab}, \citenamefont
  {Constantinou}, \citenamefont {Strocov}, \citenamefont {Usanov} \emph
  {et~al.}}]{krempasky2024altermagnetic}%
  \BibitemOpen
  \bibfield  {author} {\bibinfo {author} {\bibfnamefont {J.}~\bibnamefont
  {Krempask{\`y}}}, \bibinfo {author} {\bibfnamefont {L.}~\bibnamefont
  {{\v{S}}mejkal}}, \bibinfo {author} {\bibfnamefont {S.}~\bibnamefont
  {D’Souza}}, \bibinfo {author} {\bibfnamefont {M.}~\bibnamefont {Hajlaoui}},
  \bibinfo {author} {\bibfnamefont {G.}~\bibnamefont {Springholz}}, \bibinfo
  {author} {\bibfnamefont {K.}~\bibnamefont {Uhl{\'\i}{\v{r}}ov{\'a}}},
  \bibinfo {author} {\bibfnamefont {F.}~\bibnamefont {Alarab}}, \bibinfo
  {author} {\bibfnamefont {P.}~\bibnamefont {Constantinou}}, \bibinfo {author}
  {\bibfnamefont {V.}~\bibnamefont {Strocov}}, \bibinfo {author} {\bibfnamefont
  {D.}~\bibnamefont {Usanov}}, \emph {et~al.},\ }\bibfield  {title} {\bibinfo
  {title} {Altermagnetic lifting of kramers spin degeneracy},\ }\href@noop {}
  {\bibfield  {journal} {\bibinfo  {journal} {Nature}\ }\textbf {\bibinfo
  {volume} {626}},\ \bibinfo {pages} {517} (\bibinfo {year}
  {2024})}\BibitemShut {NoStop}%
\bibitem [{\citenamefont {Zhu}\ \emph {et~al.}(2024)\citenamefont {Zhu},
  \citenamefont {Chen}, \citenamefont {Liu}, \citenamefont {Liu}, \citenamefont
  {Liu}, \citenamefont {Zha}, \citenamefont {Qu}, \citenamefont {Hong},
  \citenamefont {Li}, \citenamefont {Jiang} \emph
  {et~al.}}]{zhu2024observation}%
  \BibitemOpen
  \bibfield  {author} {\bibinfo {author} {\bibfnamefont {Y.-P.}\ \bibnamefont
  {Zhu}}, \bibinfo {author} {\bibfnamefont {X.}~\bibnamefont {Chen}}, \bibinfo
  {author} {\bibfnamefont {X.-R.}\ \bibnamefont {Liu}}, \bibinfo {author}
  {\bibfnamefont {Y.}~\bibnamefont {Liu}}, \bibinfo {author} {\bibfnamefont
  {P.}~\bibnamefont {Liu}}, \bibinfo {author} {\bibfnamefont {H.}~\bibnamefont
  {Zha}}, \bibinfo {author} {\bibfnamefont {G.}~\bibnamefont {Qu}}, \bibinfo
  {author} {\bibfnamefont {C.}~\bibnamefont {Hong}}, \bibinfo {author}
  {\bibfnamefont {J.}~\bibnamefont {Li}}, \bibinfo {author} {\bibfnamefont
  {Z.}~\bibnamefont {Jiang}}, \emph {et~al.},\ }\bibfield  {title} {\bibinfo
  {title} {Observation of plaid-like spin splitting in a noncoplanar
  antiferromagnet},\ }\href@noop {} {\bibfield  {journal} {\bibinfo  {journal}
  {Nature}\ }\textbf {\bibinfo {volume} {626}},\ \bibinfo {pages} {523}
  (\bibinfo {year} {2024})}\BibitemShut {NoStop}%
\bibitem [{\citenamefont {Das}\ \emph {et~al.}(2012)\citenamefont {Das},
  \citenamefont {Ronen}, \citenamefont {Most}, \citenamefont {Oreg},
  \citenamefont {Heiblum},\ and\ \citenamefont {Shtrikman}}]{das2012zero}%
  \BibitemOpen
  \bibfield  {author} {\bibinfo {author} {\bibfnamefont {A.}~\bibnamefont
  {Das}}, \bibinfo {author} {\bibfnamefont {Y.}~\bibnamefont {Ronen}}, \bibinfo
  {author} {\bibfnamefont {Y.}~\bibnamefont {Most}}, \bibinfo {author}
  {\bibfnamefont {Y.}~\bibnamefont {Oreg}}, \bibinfo {author} {\bibfnamefont
  {M.}~\bibnamefont {Heiblum}},\ and\ \bibinfo {author} {\bibfnamefont
  {H.}~\bibnamefont {Shtrikman}},\ }\bibfield  {title} {\bibinfo {title}
  {Zero-bias peaks and splitting in an al--inas nanowire topological
  superconductor as a signature of majorana fermions},\ }\href@noop {}
  {\bibfield  {journal} {\bibinfo  {journal} {Nature Physics}\ }\textbf
  {\bibinfo {volume} {8}},\ \bibinfo {pages} {887} (\bibinfo {year}
  {2012})}\BibitemShut {NoStop}%
\bibitem [{\citenamefont {Pozo}\ \emph {et~al.}(2019)\citenamefont {Pozo},
  \citenamefont {Repellin},\ and\ \citenamefont
  {Grushin}}]{PhysRevLett.123.247401}%
  \BibitemOpen
  \bibfield  {author} {\bibinfo {author} {\bibfnamefont {O.}~\bibnamefont
  {Pozo}}, \bibinfo {author} {\bibfnamefont {C.}~\bibnamefont {Repellin}},\
  and\ \bibinfo {author} {\bibfnamefont {A.~G.}\ \bibnamefont {Grushin}},\
  }\bibfield  {title} {\bibinfo {title} {Quantization in chiral higher order
  topological insulators: Circular dichroism and local chern marker},\ }\href
  {https://doi.org/10.1103/PhysRevLett.123.247401} {\bibfield  {journal}
  {\bibinfo  {journal} {Phys. Rev. Lett.}\ }\textbf {\bibinfo {volume} {123}},\
  \bibinfo {pages} {247401} (\bibinfo {year} {2019})}\BibitemShut {NoStop}%
\bibitem [{\citenamefont {Pan}\ \emph {et~al.}(2022)\citenamefont {Pan},
  \citenamefont {Li}, \citenamefont {Fan},\ and\ \citenamefont
  {Huang}}]{pan2022two}%
  \BibitemOpen
  \bibfield  {author} {\bibinfo {author} {\bibfnamefont {M.}~\bibnamefont
  {Pan}}, \bibinfo {author} {\bibfnamefont {D.}~\bibnamefont {Li}}, \bibinfo
  {author} {\bibfnamefont {J.}~\bibnamefont {Fan}},\ and\ \bibinfo {author}
  {\bibfnamefont {H.}~\bibnamefont {Huang}},\ }\bibfield  {title} {\bibinfo
  {title} {Two-dimensional stiefel-whitney insulators in liganded xenes},\
  }\href@noop {} {\bibfield  {journal} {\bibinfo  {journal} {npj Computational
  Materials}\ }\textbf {\bibinfo {volume} {8}},\ \bibinfo {pages} {1} (\bibinfo
  {year} {2022})}\BibitemShut {NoStop}%
\bibitem [{\citenamefont {Guo}\ \emph {et~al.}(2022)\citenamefont {Guo},
  \citenamefont {Deng}, \citenamefont {Xie},\ and\ \citenamefont
  {Wang}}]{guo2022quadrupole}%
  \BibitemOpen
  \bibfield  {author} {\bibinfo {author} {\bibfnamefont {Z.}~\bibnamefont
  {Guo}}, \bibinfo {author} {\bibfnamefont {J.}~\bibnamefont {Deng}}, \bibinfo
  {author} {\bibfnamefont {Y.}~\bibnamefont {Xie}},\ and\ \bibinfo {author}
  {\bibfnamefont {Z.}~\bibnamefont {Wang}},\ }\bibfield  {title} {\bibinfo
  {title} {Quadrupole topological insulators in ta2m3te5 (m= ni, pd)
  monolayers},\ }\href@noop {} {\bibfield  {journal} {\bibinfo  {journal} {npj
  Quantum Materials}\ }\textbf {\bibinfo {volume} {7}},\ \bibinfo {pages} {87}
  (\bibinfo {year} {2022})}\BibitemShut {NoStop}%
\bibitem [{\citenamefont {Rauch}\ \emph {et~al.}(2018)\citenamefont {Rauch},
  \citenamefont {Olsen}, \citenamefont {Vanderbilt},\ and\ \citenamefont
  {Souza}}]{PhysRevB.98.115108}%
  \BibitemOpen
  \bibfield  {author} {\bibinfo {author} {\bibfnamefont {T.~c.~v.}\
  \bibnamefont {Rauch}}, \bibinfo {author} {\bibfnamefont {T.}~\bibnamefont
  {Olsen}}, \bibinfo {author} {\bibfnamefont {D.}~\bibnamefont {Vanderbilt}},\
  and\ \bibinfo {author} {\bibfnamefont {I.}~\bibnamefont {Souza}},\ }\bibfield
   {title} {\bibinfo {title} {Geometric and nongeometric contributions to the
  surface anomalous hall conductivity},\ }\href
  {https://doi.org/10.1103/PhysRevB.98.115108} {\bibfield  {journal} {\bibinfo
  {journal} {Phys. Rev. B}\ }\textbf {\bibinfo {volume} {98}},\ \bibinfo
  {pages} {115108} (\bibinfo {year} {2018})}\BibitemShut {NoStop}%
\bibitem [{\citenamefont {Chen}\ \emph {et~al.}(2021)\citenamefont {Chen},
  \citenamefont {Li}, \citenamefont {Sun}, \citenamefont {Liu}, \citenamefont
  {Zhao}, \citenamefont {Lu},\ and\ \citenamefont
  {Xie}}]{PhysRevB.103.L241409}%
  \BibitemOpen
  \bibfield  {author} {\bibinfo {author} {\bibfnamefont {R.}~\bibnamefont
  {Chen}}, \bibinfo {author} {\bibfnamefont {S.}~\bibnamefont {Li}}, \bibinfo
  {author} {\bibfnamefont {H.-P.}\ \bibnamefont {Sun}}, \bibinfo {author}
  {\bibfnamefont {Q.}~\bibnamefont {Liu}}, \bibinfo {author} {\bibfnamefont
  {Y.}~\bibnamefont {Zhao}}, \bibinfo {author} {\bibfnamefont {H.-Z.}\
  \bibnamefont {Lu}},\ and\ \bibinfo {author} {\bibfnamefont {X.~C.}\
  \bibnamefont {Xie}},\ }\bibfield  {title} {\bibinfo {title} {Using nonlocal
  surface transport to identify the axion insulator},\ }\href
  {https://doi.org/10.1103/PhysRevB.103.L241409} {\bibfield  {journal}
  {\bibinfo  {journal} {Phys. Rev. B}\ }\textbf {\bibinfo {volume} {103}},\
  \bibinfo {pages} {L241409} (\bibinfo {year} {2021})}\BibitemShut {NoStop}%
\bibitem [{\citenamefont {Liu}\ \emph {et~al.}(2010)\citenamefont {Liu},
  \citenamefont {Qi}, \citenamefont {Zhang}, \citenamefont {Dai}, \citenamefont
  {Fang},\ and\ \citenamefont {Zhang}}]{PhysRevB.82.045122}%
  \BibitemOpen
  \bibfield  {author} {\bibinfo {author} {\bibfnamefont {C.-X.}\ \bibnamefont
  {Liu}}, \bibinfo {author} {\bibfnamefont {X.-L.}\ \bibnamefont {Qi}},
  \bibinfo {author} {\bibfnamefont {H.}~\bibnamefont {Zhang}}, \bibinfo
  {author} {\bibfnamefont {X.}~\bibnamefont {Dai}}, \bibinfo {author}
  {\bibfnamefont {Z.}~\bibnamefont {Fang}},\ and\ \bibinfo {author}
  {\bibfnamefont {S.-C.}\ \bibnamefont {Zhang}},\ }\bibfield  {title} {\bibinfo
  {title} {Model hamiltonian for topological insulators},\ }\href
  {https://doi.org/10.1103/PhysRevB.82.045122} {\bibfield  {journal} {\bibinfo
  {journal} {Phys. Rev. B}\ }\textbf {\bibinfo {volume} {82}},\ \bibinfo
  {pages} {045122} (\bibinfo {year} {2010})}\BibitemShut {NoStop}%
\bibitem [{\citenamefont {Varnava}\ \emph {et~al.}(2022)\citenamefont
  {Varnava}, \citenamefont {Berry}, \citenamefont {McQueen},\ and\
  \citenamefont {Vanderbilt}}]{varnava2022engineering}%
  \BibitemOpen
  \bibfield  {author} {\bibinfo {author} {\bibfnamefont {N.}~\bibnamefont
  {Varnava}}, \bibinfo {author} {\bibfnamefont {T.}~\bibnamefont {Berry}},
  \bibinfo {author} {\bibfnamefont {T.~M.}\ \bibnamefont {McQueen}},\ and\
  \bibinfo {author} {\bibfnamefont {D.}~\bibnamefont {Vanderbilt}},\ }\bibfield
   {title} {\bibinfo {title} {Engineering magnetic topological insulators in eu
  5 m 2 x 6 zintl compounds},\ }\href@noop {} {\bibfield  {journal} {\bibinfo
  {journal} {Physical Review B}\ }\textbf {\bibinfo {volume} {105}},\ \bibinfo
  {pages} {235128} (\bibinfo {year} {2022})}\BibitemShut {NoStop}%
\bibitem [{\citenamefont {Fu}\ and\ \citenamefont
  {Kane}(2008)}]{PhysRevLett.100.096407}%
  \BibitemOpen
  \bibfield  {author} {\bibinfo {author} {\bibfnamefont {L.}~\bibnamefont
  {Fu}}\ and\ \bibinfo {author} {\bibfnamefont {C.~L.}\ \bibnamefont {Kane}},\
  }\bibfield  {title} {\bibinfo {title} {Superconducting proximity effect and
  majorana fermions at the surface of a topological insulator},\ }\href
  {https://doi.org/10.1103/PhysRevLett.100.096407} {\bibfield  {journal}
  {\bibinfo  {journal} {Phys. Rev. Lett.}\ }\textbf {\bibinfo {volume} {100}},\
  \bibinfo {pages} {096407} (\bibinfo {year} {2008})}\BibitemShut {NoStop}%
\bibitem [{\citenamefont {Ghorashi}\ \emph {et~al.}(2023)\citenamefont
  {Ghorashi}, \citenamefont {Hughes},\ and\ \citenamefont
  {Cano}}]{ghorashi2023altermagnetic}%
  \BibitemOpen
  \bibfield  {author} {\bibinfo {author} {\bibfnamefont {S.~A.~A.}\
  \bibnamefont {Ghorashi}}, \bibinfo {author} {\bibfnamefont {T.~L.}\
  \bibnamefont {Hughes}},\ and\ \bibinfo {author} {\bibfnamefont
  {J.}~\bibnamefont {Cano}},\ }\href@noop {} {\bibinfo {title} {Altermagnetic
  routes to majorana modes in zero net magnetization}} (\bibinfo {year}
  {2023}),\ \Eprint {https://arxiv.org/abs/2306.09413} {arXiv:2306.09413
  [cond-mat.mes-hall]} \BibitemShut {NoStop}%
\bibitem [{\citenamefont {Lutchyn}\ \emph {et~al.}(2010)\citenamefont
  {Lutchyn}, \citenamefont {Sau},\ and\ \citenamefont
  {Das~Sarma}}]{Lutchyn2010}%
  \BibitemOpen
  \bibfield  {author} {\bibinfo {author} {\bibfnamefont {R.~M.}\ \bibnamefont
  {Lutchyn}}, \bibinfo {author} {\bibfnamefont {J.~D.}\ \bibnamefont {Sau}},\
  and\ \bibinfo {author} {\bibfnamefont {S.}~\bibnamefont {Das~Sarma}},\
  }\bibfield  {title} {\bibinfo {title} {Majorana fermions and a topological
  phase transition in semiconductor-superconductor heterostructures},\ }\href
  {https://doi.org/10.1103/PhysRevLett.105.077001} {\bibfield  {journal}
  {\bibinfo  {journal} {Phys. Rev. Lett.}\ }\textbf {\bibinfo {volume} {105}},\
  \bibinfo {pages} {077001} (\bibinfo {year} {2010})}\BibitemShut {NoStop}%
\bibitem [{\citenamefont {Oreg}\ \emph {et~al.}(2010)\citenamefont {Oreg},
  \citenamefont {Refael},\ and\ \citenamefont {von Oppen}}]{Oreg2010}%
  \BibitemOpen
  \bibfield  {author} {\bibinfo {author} {\bibfnamefont {Y.}~\bibnamefont
  {Oreg}}, \bibinfo {author} {\bibfnamefont {G.}~\bibnamefont {Refael}},\ and\
  \bibinfo {author} {\bibfnamefont {F.}~\bibnamefont {von Oppen}},\ }\bibfield
  {title} {\bibinfo {title} {Helical liquids and majorana bound states in
  quantum wires},\ }\href {https://doi.org/10.1103/PhysRevLett.105.177002}
  {\bibfield  {journal} {\bibinfo  {journal} {Phys. Rev. Lett.}\ }\textbf
  {\bibinfo {volume} {105}},\ \bibinfo {pages} {177002} (\bibinfo {year}
  {2010})}\BibitemShut {NoStop}%
\bibitem [{\citenamefont {Cook}\ and\ \citenamefont
  {Franz}(2011)}]{PhysRevB.84.201105}%
  \BibitemOpen
  \bibfield  {author} {\bibinfo {author} {\bibfnamefont {A.}~\bibnamefont
  {Cook}}\ and\ \bibinfo {author} {\bibfnamefont {M.}~\bibnamefont {Franz}},\
  }\bibfield  {title} {\bibinfo {title} {Majorana fermions in a
  topological-insulator nanowire proximity-coupled to an $s$-wave
  superconductor},\ }\href {https://doi.org/10.1103/PhysRevB.84.201105}
  {\bibfield  {journal} {\bibinfo  {journal} {Phys. Rev. B}\ }\textbf {\bibinfo
  {volume} {84}},\ \bibinfo {pages} {201105} (\bibinfo {year}
  {2011})}\BibitemShut {NoStop}%
\bibitem [{\citenamefont {Alicea}(2012)}]{alicea2012new}%
  \BibitemOpen
  \bibfield  {author} {\bibinfo {author} {\bibfnamefont {J.}~\bibnamefont
  {Alicea}},\ }\bibfield  {title} {\bibinfo {title} {New directions in the
  pursuit of majorana fermions in solid state systems},\ }\href@noop {}
  {\bibfield  {journal} {\bibinfo  {journal} {Reports on progress in physics}\
  }\textbf {\bibinfo {volume} {75}},\ \bibinfo {pages} {076501} (\bibinfo
  {year} {2012})}\BibitemShut {NoStop}%
\bibitem [{\citenamefont {Akhmerov}\ \emph {et~al.}(2009)\citenamefont
  {Akhmerov}, \citenamefont {Nilsson},\ and\ \citenamefont
  {Beenakker}}]{PhysRevLett.102.216404}%
  \BibitemOpen
  \bibfield  {author} {\bibinfo {author} {\bibfnamefont {A.~R.}\ \bibnamefont
  {Akhmerov}}, \bibinfo {author} {\bibfnamefont {J.}~\bibnamefont {Nilsson}},\
  and\ \bibinfo {author} {\bibfnamefont {C.~W.~J.}\ \bibnamefont {Beenakker}},\
  }\bibfield  {title} {\bibinfo {title} {Electrically detected interferometry
  of majorana fermions in a topological insulator},\ }\href
  {https://doi.org/10.1103/PhysRevLett.102.216404} {\bibfield  {journal}
  {\bibinfo  {journal} {Phys. Rev. Lett.}\ }\textbf {\bibinfo {volume} {102}},\
  \bibinfo {pages} {216404} (\bibinfo {year} {2009})}\BibitemShut {NoStop}%
\bibitem [{\citenamefont {Wang}\ \emph {et~al.}(2015)\citenamefont {Wang},
  \citenamefont {Zhou}, \citenamefont {Lian},\ and\ \citenamefont
  {Zhang}}]{PhysRevB.92.064520}%
  \BibitemOpen
  \bibfield  {author} {\bibinfo {author} {\bibfnamefont {J.}~\bibnamefont
  {Wang}}, \bibinfo {author} {\bibfnamefont {Q.}~\bibnamefont {Zhou}}, \bibinfo
  {author} {\bibfnamefont {B.}~\bibnamefont {Lian}},\ and\ \bibinfo {author}
  {\bibfnamefont {S.-C.}\ \bibnamefont {Zhang}},\ }\bibfield  {title} {\bibinfo
  {title} {Chiral topological superconductor and half-integer conductance
  plateau from quantum anomalous hall plateau transition},\ }\href
  {https://doi.org/10.1103/PhysRevB.92.064520} {\bibfield  {journal} {\bibinfo
  {journal} {Phys. Rev. B}\ }\textbf {\bibinfo {volume} {92}},\ \bibinfo
  {pages} {064520} (\bibinfo {year} {2015})}\BibitemShut {NoStop}%
\bibitem [{\citenamefont {Fu}\ and\ \citenamefont {Kane}(2009)}]{Fu2009}%
  \BibitemOpen
  \bibfield  {author} {\bibinfo {author} {\bibfnamefont {L.}~\bibnamefont
  {Fu}}\ and\ \bibinfo {author} {\bibfnamefont {C.~L.}\ \bibnamefont {Kane}},\
  }\bibfield  {title} {\bibinfo {title} {Probing neutral majorana fermion edge
  modes with charge transport},\ }\href
  {https://doi.org/10.1103/PhysRevLett.102.216403} {\bibfield  {journal}
  {\bibinfo  {journal} {Phys. Rev. Lett.}\ }\textbf {\bibinfo {volume} {102}},\
  \bibinfo {pages} {216403} (\bibinfo {year} {2009})}\BibitemShut {NoStop}%
\bibitem [{\citenamefont {Kresse}\ and\ \citenamefont
  {Furthm\"uller}(1996)}]{PhysRevB.54.11169}%
  \BibitemOpen
  \bibfield  {author} {\bibinfo {author} {\bibfnamefont {G.}~\bibnamefont
  {Kresse}}\ and\ \bibinfo {author} {\bibfnamefont {J.}~\bibnamefont
  {Furthm\"uller}},\ }\bibfield  {title} {\bibinfo {title} {Efficient iterative
  schemes for ab initio total-energy calculations using a plane-wave basis
  set},\ }\href {https://doi.org/10.1103/PhysRevB.54.11169} {\bibfield
  {journal} {\bibinfo  {journal} {Phys. Rev. B}\ }\textbf {\bibinfo {volume}
  {54}},\ \bibinfo {pages} {11169} (\bibinfo {year} {1996})}\BibitemShut
  {NoStop}%
\bibitem [{\citenamefont {Kresse}\ and\ \citenamefont
  {Joubert}(1999)}]{PhysRevB.59.1758}%
  \BibitemOpen
  \bibfield  {author} {\bibinfo {author} {\bibfnamefont {G.}~\bibnamefont
  {Kresse}}\ and\ \bibinfo {author} {\bibfnamefont {D.}~\bibnamefont
  {Joubert}},\ }\bibfield  {title} {\bibinfo {title} {From ultrasoft
  pseudopotentials to the projector augmented-wave method},\ }\href
  {https://doi.org/10.1103/PhysRevB.59.1758} {\bibfield  {journal} {\bibinfo
  {journal} {Phys. Rev. B}\ }\textbf {\bibinfo {volume} {59}},\ \bibinfo
  {pages} {1758} (\bibinfo {year} {1999})}\BibitemShut {NoStop}%
\bibitem [{\citenamefont {Bl\"ochl}(1994)}]{PhysRevB.50.17953}%
  \BibitemOpen
  \bibfield  {author} {\bibinfo {author} {\bibfnamefont {P.~E.}\ \bibnamefont
  {Bl\"ochl}},\ }\bibfield  {title} {\bibinfo {title} {Projector augmented-wave
  method},\ }\href {https://doi.org/10.1103/PhysRevB.50.17953} {\bibfield
  {journal} {\bibinfo  {journal} {Phys. Rev. B}\ }\textbf {\bibinfo {volume}
  {50}},\ \bibinfo {pages} {17953} (\bibinfo {year} {1994})}\BibitemShut
  {NoStop}%
\bibitem [{\citenamefont {Perdew}\ \emph {et~al.}(1996)\citenamefont {Perdew},
  \citenamefont {Burke},\ and\ \citenamefont
  {Ernzerhof}}]{PhysRevLett.77.3865}%
  \BibitemOpen
  \bibfield  {author} {\bibinfo {author} {\bibfnamefont {J.~P.}\ \bibnamefont
  {Perdew}}, \bibinfo {author} {\bibfnamefont {K.}~\bibnamefont {Burke}},\ and\
  \bibinfo {author} {\bibfnamefont {M.}~\bibnamefont {Ernzerhof}},\ }\bibfield
  {title} {\bibinfo {title} {Generalized gradient approximation made simple},\
  }\href {https://doi.org/10.1103/PhysRevLett.77.3865} {\bibfield  {journal}
  {\bibinfo  {journal} {Phys. Rev. Lett.}\ }\textbf {\bibinfo {volume} {77}},\
  \bibinfo {pages} {3865} (\bibinfo {year} {1996})}\BibitemShut {NoStop}%
\bibitem [{\citenamefont {Dudarev}\ \emph {et~al.}(1998)\citenamefont
  {Dudarev}, \citenamefont {Botton}, \citenamefont {Savrasov}, \citenamefont
  {Humphreys},\ and\ \citenamefont {Sutton}}]{PhysRevB.57.1505}%
  \BibitemOpen
  \bibfield  {author} {\bibinfo {author} {\bibfnamefont {S.~L.}\ \bibnamefont
  {Dudarev}}, \bibinfo {author} {\bibfnamefont {G.~A.}\ \bibnamefont {Botton}},
  \bibinfo {author} {\bibfnamefont {S.~Y.}\ \bibnamefont {Savrasov}}, \bibinfo
  {author} {\bibfnamefont {C.~J.}\ \bibnamefont {Humphreys}},\ and\ \bibinfo
  {author} {\bibfnamefont {A.~P.}\ \bibnamefont {Sutton}},\ }\bibfield  {title}
  {\bibinfo {title} {Electron-energy-loss spectra and the structural stability
  of nickel oxide: An lsda+u study},\ }\href
  {https://doi.org/10.1103/PhysRevB.57.1505} {\bibfield  {journal} {\bibinfo
  {journal} {Phys. Rev. B}\ }\textbf {\bibinfo {volume} {57}},\ \bibinfo
  {pages} {1505} (\bibinfo {year} {1998})}\BibitemShut {NoStop}%
\bibitem [{\citenamefont {Souza}\ \emph {et~al.}(2001)\citenamefont {Souza},
  \citenamefont {Marzari},\ and\ \citenamefont
  {Vanderbilt}}]{PhysRevB.65.035109}%
  \BibitemOpen
  \bibfield  {author} {\bibinfo {author} {\bibfnamefont {I.}~\bibnamefont
  {Souza}}, \bibinfo {author} {\bibfnamefont {N.}~\bibnamefont {Marzari}},\
  and\ \bibinfo {author} {\bibfnamefont {D.}~\bibnamefont {Vanderbilt}},\
  }\bibfield  {title} {\bibinfo {title} {Maximally localized wannier functions
  for entangled energy bands},\ }\href
  {https://doi.org/10.1103/PhysRevB.65.035109} {\bibfield  {journal} {\bibinfo
  {journal} {Phys. Rev. B}\ }\textbf {\bibinfo {volume} {65}},\ \bibinfo
  {pages} {035109} (\bibinfo {year} {2001})}\BibitemShut {NoStop}%
\end{thebibliography}

%

\end{document}